\documentclass[10pt,preprint]{aastex}
\newcommand*{\BH}{{\bullet}}
\newcommand*{\dN}{{\dot N}}
\newcommand*{\eg}{{\em e.g.}}
\newcommand*{\ie}{{\em i.e.}}
\begin{document}

\defcitealias{Rajagopal95}{RR}

\title{Gravitational Waves Probe the Coalescence Rate of Massive Black Hole Binaries}

\newcommand{\cfpa}{Center for Particle
  Astrophysics, University of California, Berkeley, CA, USA}
\newcommand{\spacescience}{Space Sciences Laboratory, University of
  California, Berkeley, CA, USA} 
\newcommand{\ucbphysics}{Department of Physics, University of California,
  Berkeley CA, USA}
\newcommand{\ucbastro}{Department of Astronomy, University of California,
  Berkeley CA, USA 94720-3411}
\author{A.~H.~Jaffe}
\affil{Astrophysics Group, Blackett Laboratory, Imperial College London, SW7 2AZ ENGLAND}
\author{D.~C.~Backer}
\affil{\ucbastro}

\begin{abstract}
  We calculate the expected nHz--$\mu$Hz gravitational wave (GW) spectrum from
  coalescing Massive Black Hole (MBH) binaries resulting from mergers of
  their host galaxies. We consider detection of this spectrum by
  precision pulsar timing and a future Pulsar Timing Array. The spectrum
  depends on the merger rate of massive galaxies, the demographics of 
  MBHs at low and high redshift, and the dynamics of MBH binaries.  
  We apply recent theoretical and observational work on all of these
  fronts. The spectrum has a characteristic strain $h_c(f)\sim10^{-15}
  (f/{\rm yr}^{-1})^{-2/3}$, just below the detection limit from recent
  analysis of precision pulsar timing measurements.
  However, the amplitude of the spectrum is still very uncertain 
  owing to approximations in the theoretical formulation of the model, 
  to our lack of knowledge of the merger rate and MBH 
  population at high redshift, and to the dynamical problem of removing enough
  angular momentum from the MBH binary to reach a GW-dominated regime.
\end{abstract}


\section{Introduction}
\label{sec:intro}

Over the past decade, evidence has accumulated that Massive Black 
Holes\footnote{The literature has variable usage of the terms ``Massive'' and
``Supermassive'' to describe the $10^{6\mbox{--}9}$ M$_\odot$ black holes
in the centers of galaxies \citep[\eg,][]{HughesSnowmass2001}. For
concision, we favor the former in this work.}
(MBHs) are ubiquitous in the spheroidal components of low-redshift
galaxies \citep[\eg,][]{mag98}, and that their masses are tightly
correlated with the properties of the host galaxies. These initial
studies noted that the MBH mass scaled with the mass 
of the spheroidal component of the parent galaxy
as inferred from the luminosity 
although there was considerable scatter.
More recently, a much tighter correlation has been observed between the
MBH mass and the velocity dispersion in the 
spheroid \citep{ferrarese,gebhardt}.

At higher redshift, evidence for the ubiquity of MBHs is more
circumstantial. However, we do know that many distant galaxies harbor active
nuclei for some portion of their life and, in turn, that
the engine driving this activity
is almost certainly gravitational accretion onto a central black hole.
There are many outstanding questions regarding 
the era and duty cycle of AGN activity, 
the initial mass and mass evolution of MBHs, 
and the mass growth of MBHs during AGN episodes.
Nonetheless, the existing evidence strongly favors the hypothesis
that the present-day MBH population is the dormant remnant 
of the AGN population at high redshift.

During the last decade astrophysicists have also begun to
understand the important role that mergers play in the history of
galaxies. In the hierarchical clustering picture, mergers are by
definition responsible for building up the mass of today's galaxies. 
Numerical simulations have also shown that the merger history also
determines the morphology of present-day galaxies: \eg, major mergers 
tend to produce elliptical galaxies. The strongest evidence for the
mergers is the relatively flat density scaling laws in ``some''
spheroids \citep{MilMer01}.
Simulations that reproduce this scaling actually
calculate the merger of galaxy halos where the majority
of the mass resides.  The subsequent merger of the central galaxies
within the halos is nearly guaranteed owing to the rapid process of
dynamical friction for these dynamically soft stellar systems.

The ubiquity of MBHs in galaxies and the frequent growth of galaxies by
merger leads to the obvious question about the fate of the pair of MBHs
in the subsequent evolution of merged systems.  The same dynamical
friction process that brings together parent galaxies within merged
halos will also lead to the sinking of the MBHs to the center of the
merger daughter galaxy.  Unfortunately, in the simplest case dynamical
friction will ``turn off'' well before the MBH binary
coalesces\footnote{We use the term {\it merger} to denote the joining of
  halo and galaxy pairs and the term {\it coalescence} for the joining
  of the members of a MBH binary to distinguish implicitly the two
  nominally sequential processes.}, or even before it reaches a regime
where gravitational radiation losses will drive the binary's evolution
\citep{Begelman80}.  Various mechanisms have been suggested to evolve
the MBH pair through this period.

If the pair can indeed evolve to the gravitational wave (GW) regime, it will
over time lose angular momentum to gravitational radiation and
eventually coalesce. The final inspirals of MBH pairs take about $10^4$
seconds, and are the most luminous gravitational wave events in the Universe.
These events are one of the chief targets of the Laser Interferometer
Space Antenna (LISA) GW
interferometer satellite. In this paper, we concentrate on
the quiescent evolution leading up to this state.  Before the final
inspiral, the GW amplitude is
\begin{equation}
  \label{eq:gwamp}
  h\simeq 4.4\times 10^{-17}M_8^{5/3}P_{\rm yr}^{-2/3}D_{\rm Gpc}^{-1}{q\over(1+q)^2},
\end{equation}
where $h$ is the strain or metric perturbation, 
$M_8$ is the total mass of the binary in units of $10^8 M_\odot$, 
$q$ is the mass ratio ($q<1$), $P_{\rm yr}$ is the observed GW period in years 
(the orbital period divided by 2 for the expected
nearly-circular orbits), and $D_{\rm Gpc}$ is the distance
in Gpc. The lifetime of the system is
\begin{equation}
  \label{eq:tGW}
  t_{\rm GW} = 1.1\times 10^6 {\rm~yr}\;M_8^{-5/3}P_{\rm yr}^{8/3}{(1+q)^2\over q}.
\end{equation}

The precision of the rotation periods of millisecond period pulsars
which is established via pulse arrival time measurements allows
detection of the stochastic MBH-MBH background spectrum at nHz
frequencies \citep{Sazhin78, Detweiler79, Rajagopal95}.  The detection
process is very similar to that of the laser interferometers---passage
of an electromagnetic signal through distorted space-time---except that
in the case of pulsars no mirrors are needed as we have a distant
precision clock, the rotating neutron star, sending us a periodic pulse
train.  Measurements of a single pulsar can place an upper limit on the
presence of a spectrum of gravitational radiation \citep{Kaspi94,
  Lommen01b}.  In short, the limit on strain is given by the ratio of
the timing precision and the measurement duration; {\it e.g.} 1
$\mu$s/10 yr $\sim~3\times 10^{-14}$. While pulsar timing cannot detect
individual events of the small amplitude given in
equation~(\ref{eq:gwamp}), these measurements have a good prospect for
detection of the stochastic background of GWs from the ensemble of these
MBH-MBH coalescence events throughout the Universe\footnote{We refer to
  this as {\it background} radiation, although those interested in {\it
    relic} radiation from the first inflationary moments might choose to
  call this {\it foreground} radiation.}.  The GW perturbs pulse arrival
times of a spatial array of pulsars in a correlated manner that is
distinct from other known perturbations such as atomic time and
ephemeris errors. A Pulsar Timing Array then acts as a nHz gravitational wave
telescope capable of direct detection of the stochastic background
radiation.

Questions remain. Are there dynamical processes that drive the
MBH pair into the GW regime? Does this happen frequently?  If so, can we
observe the gravitational radiation via pulsar timing?  If not, do we
then see evidence of close binary MBHs ``hung up'' in the centers of
many, or most, galaxies?  If we observe neither the GW signal via
precision pulsar timing nor inactive close pairs via high angular
resolution studies, does this imply that some part of the paradigm---the
ubiquity of black holes at all redshifts and the importance of mergers
in galaxy evolution---is flawed? Some of these questions have been taken
up by other groups over the years. In their important paper,
\citet[hereafter RR]{Rajagopal95}, discussed several mechanisms driving
the MBH pair into the GW regime, and calculated the GW spectrum under
various assumptions.  More recently, \citet{Gould99} follow up on the
gas-dynamical mechanism for driving the MBH coalescence first mentioned
by \cite{Begelman80}. \citet{ArmitageNatarajan02} explore the
detailed interaction of the MBH binary with the accretion disk.
Meanwhile, \citet{MilMer01} and \citet{Yu01} have
investigated the stellar-dynamical schemes. Finally, \citet{Menou01}
have investigated the effect that a time-dependent change in MBH
``demographics'' might have on the merger history of galaxies, 
and \citet{Phinney01}
has shown an alternative method for calculating the GW spectrum for
generic sources.

The advances in our understanding of MBHs along with steady
progress in precision pulsar timing have led us to this paper.
In \S\ref{sec:ingredients} we first assemble the ingredients needed to
model the stochastic GW background from MBH binaries, and in
\S\ref{sec:strainspec} we present the results of our calculations. In
\S\ref{sec:PT} we discuss the detection of gravitational radiation with
a Pulsar Timing Array and the status of current experiments.
In the concluding section we
discuss the theoretical and observational prospects for a better
understanding of the various ingredients to our calculations and
measurements.

\section{The Binary MBH Gravity Wave Spectrum}
\label{sec:ingredients}

First, we need to define the terminology for a stochastic gravitational wave
background spectrum that has been presented in the literature in a
variety of ways (\eg, \citet{Burke75,AllenRomano99,Maggiore00}).   A single
GW is denoted by the transverse-traceless part of the metric
perturbation, $h_{ab}({\bf x},t)=h_{+}(t)e^+_{ab}({\bf k}) +
h_{\times}(t)e^\times_{ab}({\bf k})$, where plus signs and crosses refer to the
two polarization degrees of freedom of a gravitational wave, and the $e^i_{ab}$
are basis tensors for the polarizations which are functions of the
direction of propagation of the gravitational wave, ${\bf k}$. The two amplitudes combine
with the two coordinate directions and the polarization position angle
to yield five independent parameters of the single-wave metric
perturbation.  The stochastic background will excite both wave
components equally and
randomly $\langle |h_+|^2 \rangle = \langle
|h_\times|^2 \rangle$ and $\langle h^*_+ h_\times\rangle=0$, and all
directions and position angles will be equally likely.

The total power spectral density of gravitational waves, $S_h(f)$, is defined by
\begin{equation}
  \label{eq:Sh}
  \sum_{P=+,\times}
  \langle {\tilde h}_P(f){\tilde h}_P^*(f') \rangle 
  =\frac{1}{2}\sum_{a,b} \langle{\tilde h}_{ab}(f){\tilde h}_{ab}^*(f')\rangle
  =\frac{1}{2} \delta(f-f') S_h(f),
\end{equation}
where ${\tilde h}_P(f)$ is the Fourier transform of the $P=+,\times$
polarization component of the metric strain tensor at frequency $f$, and
$\delta$ is the Dirac delta function (which comes from our Fourier
conventions and the definition of $S_h(f)$ as a spectral density with
units of inverse frequency). There are two spectrum conventions in the
literature: ``two-sided'' (defined for $|f|<\infty$) and
``one-sided'' ($f\ge0$).  They differ in amplitude by a factor of
two: $S_{\rm one}=2S_{\rm two}$. 
We use the one-sided version in this
paper.  From this, we can define other useful quantities, such as the
fractional contribution to the energy density of the Universe from a
logarithmic interval of frequency,
\begin{equation}
  \label{eq:omega}
  \Omega_{\rm GW}(f) = \frac{1}{\rho_c} \frac{d\rho_\mathrm{GW}}{d\ln f}
 = \frac{2\pi^2}{3H_0^2} f^3 S_h(f),
\end{equation}
where $H_0=100~h_{0}~$km s$^{-1}$ Mpc$^{-1}$ is Hubble's constant,
and $\rho_c=3H_0^2/8\pi G$ is the critical density in an FRW
(Freedman-Robertson-Walker) universe.
Finally, we can also write down an expression relating the power
spectrum to a ``characteristic strain'' spectrum, $h_c(f)$,
\begin{equation}
  \label{eq:hc}
  h_c(f) = \sqrt{f S_h(f)}\;.
\end{equation}
We return in \S\ref{sec:PT} to a discussion of the {\it spatial}
spectrum of the stochastic background as we formulate the response
of the Pulsar Timing Array to these waves.

In the following section, we will calculate the number density of
sources in a given interval of strain and frequency, 
$N(h,f)\;dh\;df$.\footnote{Here and throughout, 
  we use the notation $N(\cdot)$ to refer to 
the general concept of a number density function,
and not to some specific function of the arguments.
Thus, $N(z,f)\; dz\;df$
refers to the number density of binaries in a redshift and frequency interval,
  whereas $N(h,f)\; dh\; df$ refers to the number density of binaries in a
  strain and frequency interval.}
This is related to the strain power spectrum by
\begin{equation}
  \label{eq:h2N}
  S_h(f) = \int_0^\infty h_{\rm rms}^2 N(h,f)\; dh\;.
\end{equation}
Here, $h_{\rm rms}$ is defined so that the integrated power from a
single event counted by $N(h,f)$ is $\int df\; S_h(f)=h_{\rm rms}^2$.

Thus, we need to combine various ingredients to calculate the observed 
gravitational wave spectrum due to binary MBHs: the galaxy merger
rate, the black hole population demographics amongst galaxies, 
MBH binary dynamics and MBH
binary gravitational wave emission, all of which enter $N(h,f)$. In order to
calculate $N(h,f)$, we first consider a related quantity,
$\dN(z)\; dz$, the \emph{rate} of coalescence events happening in redshift
interval $dz$. Since this is a rate, the \emph{number} of events of a given
type in a time bin
$dt$ is $\dN(z)\;dz\;dt = N(h,f)\;dh\;df$.
We can relate this number to what we want by
\begin{equation}
  \label{eq:Nzt}
  N(h,f)=\dN(z) \frac{dz}{dh}\frac{dt}{df}
    = \dN(z)\frac{dz}{dh} \frac{dt}{dt_p} \frac{dt_p}{df_p} \frac{df_p}{df},
\end{equation}
where the subscript $p$ denotes proper values in the rest frame at
$r(z)$. Hence,
\begin{equation}
  N(h,f)= \dN[z(h)] \frac{dz}{dh} \left(f_p\frac{dt_p}{df_p}\right) {(1+z)\over f}
  \equiv \dN[z(h)] \frac{dz}{dh} \tau_{\rm GW} {(1+z)\over f},
\end{equation}
where we have introduced the gravitational wave timescale measured in the rest-frame: 
\begin{equation}
  \label{eq:taugwdef}
  \tau_{\rm GW}\equiv  \left(f_p\frac{dt_p}{df_p}\right).
\end{equation}
In this derivation we have assumed that the formation rate of the MBH
binaries changes slowly on cosmological timescales,
and that the binary enters the gravitational wave regime rapidly, 
so the rate is only a
function of redshift and not, say, of the initial time of the halo merger.
This should be an excellent
approximation, as argued in \citet{Phinney01}.

\subsection{Galaxy Merger/Black Hole Coalescence Rate}
\label{sec:merger}

We would therefore like to calculate $\dN(z,M_1, M_2)\; dz\; dM_1\; dM_2$,
the rate of black hole binary coalescence events observed at $z=0$
occurring in a 
given redshift interval $dz$ at $z$, due to MBHs in mass intervals $dM_1$ and
$dM_2$. Because the formation mechanism for these objects in galactic
halos is at best poorly understood, we will make a simplifying ansatz:
\begin{equation}
  \label{eq:nuz}
  \dN(z,M_1, M_2) = \nu(z) \frac{\phi(M_1,z) \phi(M_2,z)}{\phi_\BH^2},
\end{equation}
where $\nu(z)$ is the total merger rate of the spheroids in
which the MBHs reside in a notation similar to that of
\citetalias{Rajagopal95} and $\phi(M,z)$ is the mass function of 
  black holes at redshift $z$ (normalized so $\int dM
\phi(M,z)=\phi_{\BH}(z)$, the number density of MBHs at $z$).
That is, we assume that the coalescence rate of black hole binaries is
both independent of their masses and the properties of the halos in which
they are present, and rapid on cosmological timescales. 
We will see that we are primarily interested in integrals over the full
distribution, in which case the individual factors here can
be seen as appropriately weighted averages over the multivariate distribution.
We return
to the determination of the black hole mass function in \S\ref{sec:BHpop},
and proceed next with formulation of the galaxy merger rate.

We start by considering a shell of some proper depth $dl_p=cdt_p$ at
redshift $z$. This shell has a proper volume $dV_p$, or comoving
volume $dV_c$ of
\begin{equation}
  \label{eq:volelt}
  dV_p = 4\pi d_A^2~dl_p \equiv (1+z)^{-3}~dV_c,
\end{equation}
where $d_A$ is the angular diameter distance.  This distance is related
to the FRW coordinate distance, $r$, by 
$d_A = H_0^{-1}[(a_0 H_0 r)/(1+z)]$; 
note that, for example,  \citet{Peebles} refers to $r(z)$ itself as the
``angular size distance''. The FRW distance is a function of $z$ and
cosmological parameters as given by the dimensionless integral
\begin{equation}
  \label{eq:a0h0rgen}
  a_0 H_0 r(z) = \frac{a_0 H_0 {\cal R}}{c} {\cal S}_k \left[{c\over a_0 H_0 {\cal
        R}}\int_0^z
    {dz'\over E(z')}\right]\;,
\end{equation}
where $H_0=100\; h_{0}\; {\rm km}~{\rm s}^{-1} {\rm Mpc}^{-1}$ is the
Hubble constant, $a_0$ is the FRW scale factor today, $({\dot a/a})^2 =
H^2_0 E^2(z) = H_0^2\left[ \Omega_m (1+z)^3 + \Omega_\Lambda +
  (1-\Omega_m-\Omega_\Lambda)(1+z)^2\right]$, ${\dot a}=da/dt_p$, ${\cal S}_k(x)$ is
$(\sin{x}, x, \sinh{x})$ for (closed, flat, open) geometries, and ${\cal
 R}$
gives the curvature of space. In a flat Universe to which will
specialize hereafter,
\begin{equation}
  \label{eq:a0h0r}
  a_0 H_0 r(z) = \int_0^z \; { dz'\over E(z')} \qquad {\rm (flat)}.
\end{equation}

We define the merger event rate per comoving unit volume per proper unit time 
at coordinate 
time $t$ as $R(t)$.
The total event rate for the shell is $R(t)\; dV_c$. The observed rate from 
the shell requires correction of $R$ for the
redshift of the time interval, $dt_p/dt=1/(1+z)$, and then
substitution of earlier results:
\begin{equation}
  \label{eq:timerate}
  g(t) \; dt = R(t)\; dV_c\; dt_p = R(t)\; dt_p\; 4\pi c
  (1+z)^2 d_A^2 = R(t)\;dt_p\; 4\pi c^3 H_0^{-2} (a_0H_0r)^2\; .
\end{equation}
Converting the differential from time to redshift, we finally have
\begin{equation}
  \label{eq:zrate}
  \nu(z)\;dz = g(t) \;dt = 4\pi c^3 R(z) H_0^{-3} 
  {\left[a_0 H_0r(z)\right]^2 \over (1+z)E(z)} dz.
\end{equation}
That is, $\int g(t)\; dt=\int\nu(z)\;dz$ gives the rate observed at
$z=0$, with contributions from all along the past light cone.  This is
just the same as equation~(12) of \citetalias{Rajagopal95}, except that we
write our event rate as $R$ per unit proper time rather than their $F_m$ per
unit redshift, although we calculate it as a function of $z$ (so $F_m \;
dz = R\; dt_p$). Also, they specialize to a flat universe with no
cosmological constant. In particular, their equation~(14) is equivalent
to our expression with $R(z)$ equal to a constant.

\citet{Carlberg00}, following the method of \citet{Patton00}, estimate
the galaxy merger rate out to $z\simeq1$ in the CNOC2 and CFGRS redshift
surveys. They count the number of ``kinematically close pairs'',
those likely to be gravitationally infalling, based on their positions
and redshifts. They find that roughly half of ``close'' pairs observed
in projection at $20 h^{-1}$ kpc are physically close based on
morphological disturbances of the pairs. The number of pairs detected is
close to what is expected from the two-point correlation function.
Because the total number of pairs in their sample at this separation is
small, they determine the overall fraction at this separation by
extrapolating the two-point correlation function of galaxies from
measurements out to $100 h^{-1}$ kpc. Using dynamical arguments and
results of simulations, they find that the average time for a merger starting at
this separation of $20 h^{-1}$ kpc is $T=0.3$ Gyr, and that a fraction
${\cal F}\sim0.3$ of such pairs will merge in this time. Approximately,
then, ${\cal F}/T\sim 1 \;{\rm Gyr}^{-1}$ gives the fractional merger
rate for these ``close pairs,'' which should hold roughly constant as
long as the galaxy population is similar to that of the low-redshift
universe.

\begin{figure}[htbp]
\plotone{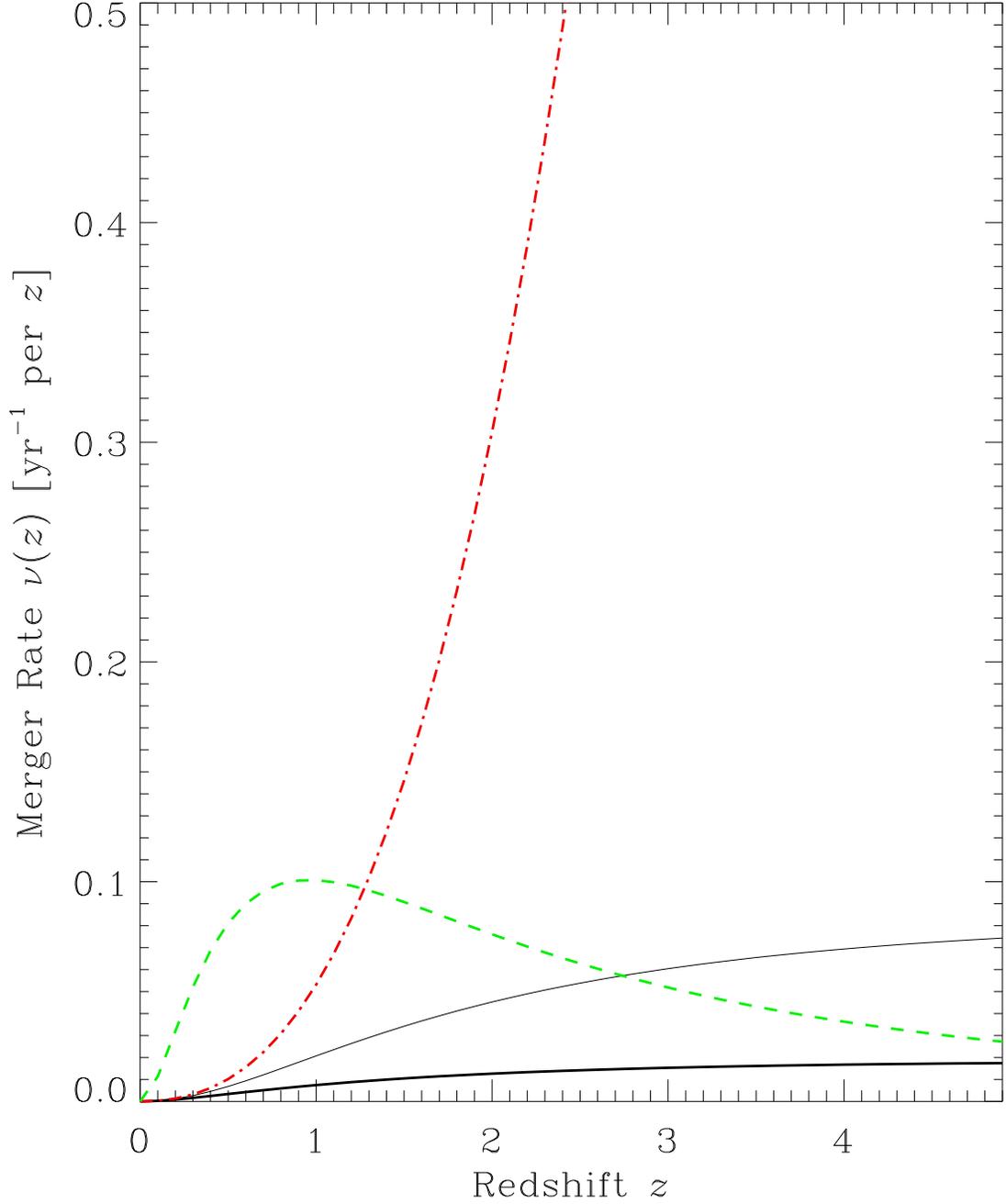}
  \caption{Various scenarios for black hole binary coalescence rates. 
    \emph{Light solid line}: equation~(\ref{eq:numrate}) (with $\Omega_m=0.3$,
    $\Omega_\Lambda=0.7$, $\gamma=2.0$); \emph{heavy solid line}:
    equation~(\ref{eq:numrate}) ($\Omega_m=1.0$,
    $\Omega_\Lambda=0.0$, $\gamma=2.0$); dash-dotted: equation~(12) from
    \citetalias{Rajagopal95}; dashed: equation~(13) from
    \citetalias{Rajagopal95}.}
  \label{fig:mergerates}
\end{figure}

Using these arguments along with observations of the number of close
pairs from the CNOC2 and CFGRS surveys over $0\le z\le 1$,
\citet{Carlberg00} find
\begin{equation}
  \label{eq:Rzcarlberg}
  R(z) \equiv \phi_\BH(z)R_g(z) 
  \simeq \phi_\BH(z) (0.09\pm0.05) (1+z)^{0.1\pm0.5} \left ({{\cal F}\over0.3}\right )
  \left ({0.3\; {\rm Gyr}\over T}\right ) \;{\rm Gyr}^{-1}, 
\end{equation}
where $R_g$ is the merger rate for a single object and $\phi_\BH(z)$ is the
comoving volume density of merging objects which for us is spheroids containing MBHs. 
We will from here on assume the merger-rate
parameters to be equal to these canonical values given in
equation~(\ref{eq:Rzcarlberg}). Crucially, we also assume that the measured
merger rate per galaxy is equal to the desired merger rate. 
Inserting this merger rate into equation~(\ref{eq:zrate}), we find
\begin{equation}
  \label{eq:numrate}
  \nu(z) \; dz = 0.03 \;{\rm yr}^{-1} 
  \left[\frac{\phi_\BH(z)}{10^{-3} h_0^3 {\rm~Mpc}^{-3}}\right]
  \left[\frac{R_0}{0.09 {\rm~Gyr}^{-1}}\right]
  \left[\frac{R_g(z)}{R_0}\right]
  \left[\frac{(a_0 H_0 r)^2}{(1+z)E(z)}\right] 
\; dz \,.
\end{equation}
We have taken $\phi_\BH=1.0\times10^{-3} h_0^3 {\rm~Mpc}^{-3}$ as our
value for the number density of merging objects; 
see \S\ref{sec:BHpop}. We have left the evolution of the
merger rate, $R_g(z)/R_0$ [with normalization $R_0=R_g(0)$], unspecified.
Whereas \citet{Carlberg00} find $R_g(z)$ approximately constant,
\citet{Patton02}, a very similar group of authors using somewhat
different data, find $R_g(z)\propto(1+z)^{1.3\mbox{--}2.3}$.
To parameterize this difference (as well as other substantial
observational disagreement on the merger rate as a function of time),
we consider several other models, generically parameterizing the merger
rate per unit time, $R_g(z)$, as a power-law in the expansion factor,
$R_g(z)=R_0 (1+z)^\gamma$. We will also allow the merger rate per unit
redshift to be a power-law, which means that $R_g(z)=R_0
E(z)(1+z)(1+z)^{\gamma'-5/2}$, where we use $\gamma'-5/2$ since that has
$\gamma'=\gamma$ when $\Omega_m=1$ and $\Omega_\Lambda=0$. 
In Figure~\ref{fig:mergerates} we show the merger rate with $\gamma=0$;
increasing this exponent would increase the merger rate at higher redshift.
Below, we will find that our limited knowledge of the merger rate is
a main contributor to the uncertainty in the final GW
spectrum. \citetalias{Rajagopal95} consider merger rates with both
higher present-day normalizations and stronger redshift evolution
(see Figure~\ref{fig:mergerates}).

\clearpage
\clearpage

We note as an aside an alternative to these phenomenological and
observational merger rates.  \citet{Menou01} have instead
considered theoretical calculations of the merger rate of galaxy halos.
They perform Monte Carlo simulations of the merger history of halos
using semi-analytic methods (so-called ``merger trees'') based on
numerical simulations and the Press-Schechter formalism and its
extensions. This allows them to vary the black hole mass function and
its relationship to the underlying halo mass function as a function of
time. In this work, however, they do not consider the detailed
gravitational wave spectrum that results from their models.

\begin{figure}[htbp]
\plotone{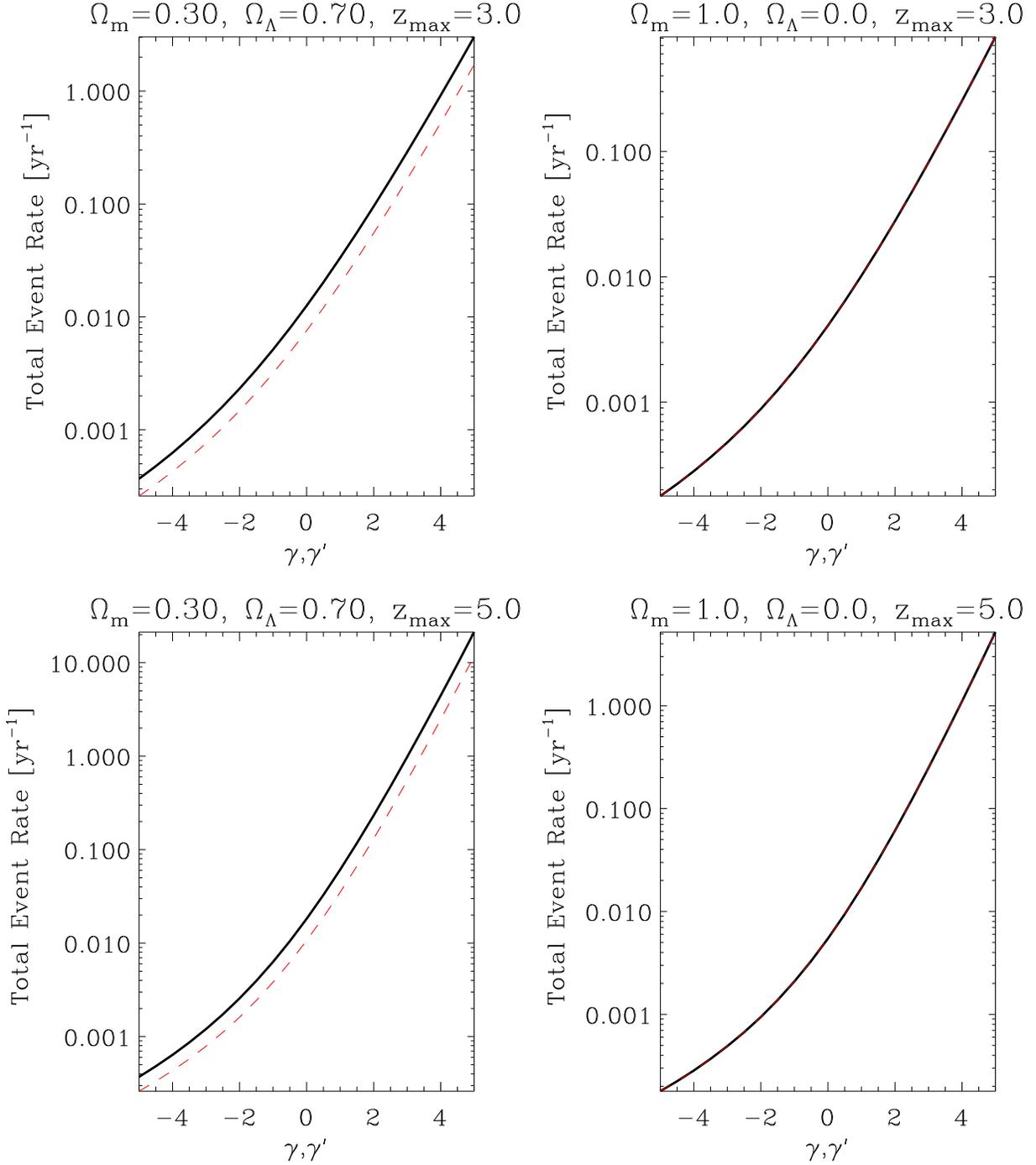}
  \caption{Total MBH binary coalescence event rate, for two sets of
    cosmological parameters and two choices of parameterizing galaxy
    merger rate, parameterized by different maximum redshifts and
    power-law indices $\gamma$ (black, solid) and $\gamma'$ (red,
    dashed) as marked. (In the right-hand panels, the curves lie atop
    one another.). We have assumed a present-day merger rate of
    $R_0=0.09\;{\rm Gyr}^{-1}$ and MBH density of
    $\phi_\BH=10^{-3}h_0^3{\rm Mpc}^{-3}$.}
  \label{fig:totalrate}
\end{figure}

Our model is useful beyond the nHz-$\mu$Hz regime. The LISA
satellite will be sensitive to the total event rate of MBH binary
inspirals as they go through their final coalescence. This event rate is
just $\int\nu(z)\;dz$, with $\nu(z)$ given by equation~(\ref{eq:zrate}) or
equation~(\ref{eq:numrate}). We show this quantity in
Figure~\ref{fig:totalrate}. For the models we have been considering, the
event rate is 0.01--1 binary coalescence events per year.
However, only a subset of these events will be detected with
sufficiently high signal-to-noise for a sufficient length of time over
the different phases of the MBH binary merger to enable a measurement of
their individual masses and distances \citep[e.g.,][]{HughesMN2001}.

\clearpage
\clearpage

\subsection{Black Hole Population Demographics}
\label{sec:BHpop}

\citet{mag98} showed that essentially all ``spheroids'' (elliptical
galaxies or the bulges of spirals) of mass $M_{\rm sph}$ have central
black of mass $M_{\BH}\sim0.006M_{\rm sph}$. More
recently~\citet{gebhardt}, \citet{ferrarese},
\citet{merritt01}
and \citet{tremaine02} have shown that the data are consistent with an
even tighter correspondence between the black hole mass and the velocity
dispersion $\sigma_v$ of the galaxy raised to a power between about 4 and 5. 
These authors find a significantly smaller ratio of $M_{\BH}$ to $M_{\rm sph}$
than that in \citet{mag98} with more careful treatment of the data.
The $\sigma_v-M_{\BH}$ relation further implies that MBHs may grow primarily 
by other process such as accretion, rather than
by the coalescence process discussed here \citep{MerFer2001}:
if the spheroids grow simply by mergers, and the MBH mass initially
tracks the spheroid mass, we might expect that the MBH mass would grow
linearly with the mass of the spheroid, whereas the velocity dispersion
raised to some power would not.
There is some controversy, which is discussed by \citet{merritt01}, 
of the implications of these results for the detailed
$M_{\BH}$--$M_{\rm sph}$ relationship. 

Determination of
either the mass function, the velocity dispersion function, 
or the luminosity function and the
mass-to-light ratio for the population of galaxies with spheroids
will provide us with the mass function of central black holes. 
Unfortunately, there have been no good measurements 
of the velocity dispersion distribution for galaxy
spheroids. Instead, we use the original, less tight, correlation between
black hole mass and spheroid mass. We will take the differential
probability  of black hole
mass at a given spheroid mass, $\phi(M_\BH)\; dM_\BH$, 
to be lognormal with mean and dispersion
given by
\begin{equation}
  \label{eq:BHdist}
  \langle \log_{10} (M_{\BH}/M_{\rm sph})\rangle = \log_{10}(0.0012) \pm 0.45
\end{equation}
\citep{MerFer2001}.
We then combine this with a theoretical model for the
spheroid luminosity function \citep[Maggorian, private
communication]{FukHogPee98,MagTre99} which is also lognormal with
\begin{equation}
  \label{eq:sphereLF}
  \langle \log_{10} (L_{\rm sph}/L_\odot)\rangle = \log_{10}(2\times10^9) \pm 0.6 
\; .
\end{equation}
Because we want an actual spatial density, we must normalize the luminosity
function with
the spheroid density, $\phi_\BH = 10^{-3}\; h_{0}^3\;{\rm
  Mpc}^{-3}$.  We also
need the spheroid mass-to-light ratio, $M/L= 7.0\;
(L/10^{10}L_\odot)^{0.2} M_\odot/L_\odot$. Putting all this together by
integrating over the two lognormal distributions gives us the MBH mass
function, also in lognormal form, with
\begin{equation}
  \label{eq:BHMF}
    \langle \log_{10} (M_{\BH}/M_\odot)\rangle = \log_{10}(1.2\times10^7) \pm 0.6
\end{equation}
This expression is appropriate for elliptical and S0 galaxies, and for
the spheroidal component of disks and spirals. There is also evidence
\citep[\eg,][]{MclureDunlop01,MclureDunlop02} that the same or very
similar relations hold for low-redshift ($z<0.2$) AGNs (Seyferts and
QSOs), although there is some dispute regarding the constants of
proportionality.

\begin{figure}[htbp]
\plotone{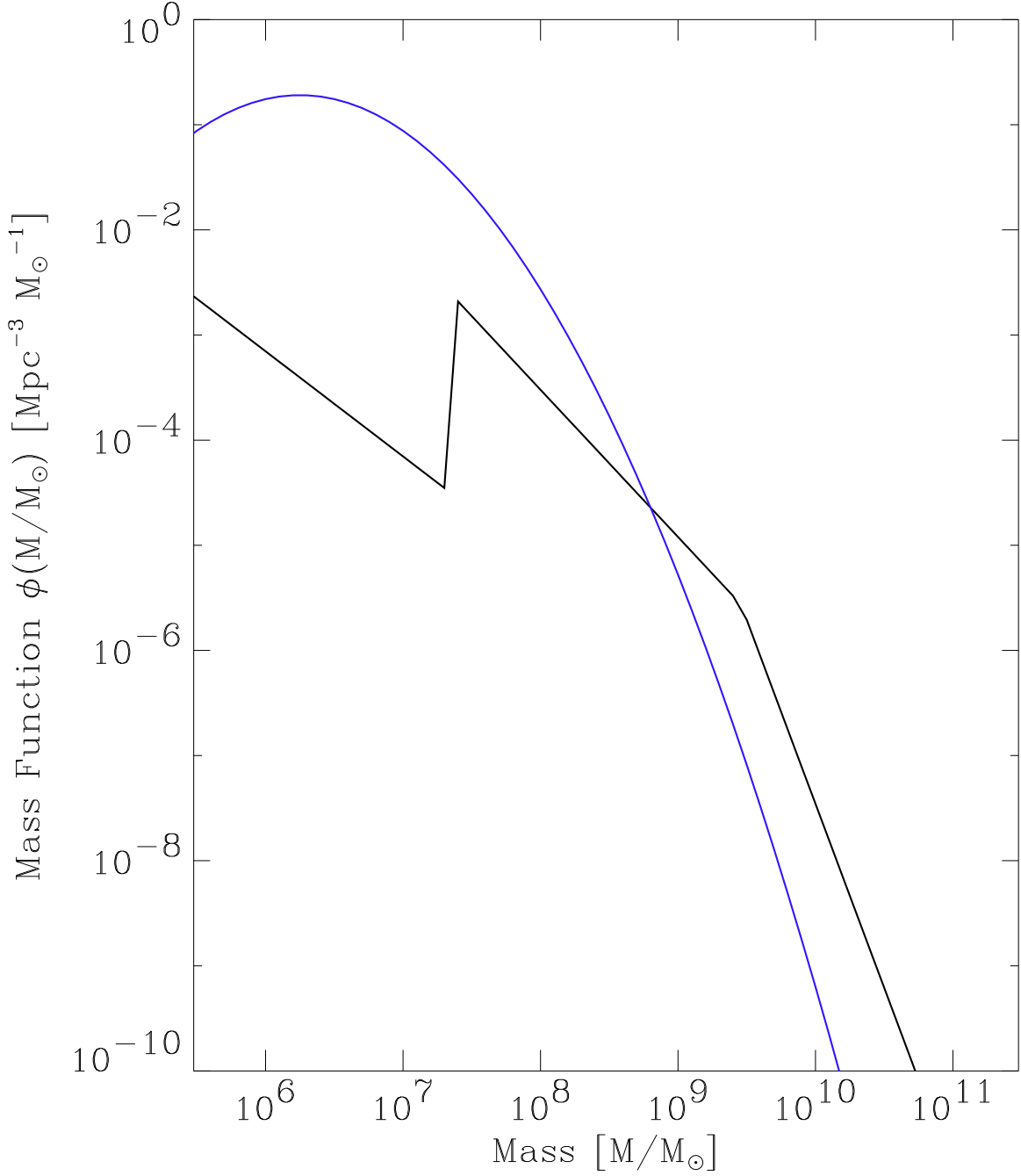}    
  \caption{The black hole mass function; the smooth upper
    curve is the lognormal model presented in the text; the black
    jagged lower curve is from \citet{SmallBlandford92}, as used in
    \citet{Rajagopal95}.}  
  \label{fig:BHmassfn}
\end{figure}

In Figure~\ref{fig:BHmassfn} we compare this mass function 
with the earlier model of \citet{SmallBlandford92}, which was
used in \citetalias{Rajagopal95}. Although the shapes of the
distributions are similar, the more recent results on demographics
predict that MBHs are more common: there are roughly an order
of magnitude more at masses below $10^9 M_\sun$.

This hard data on central black hole populations is of
necessity at low redshift. In this work we are also concerned with the
{\em evolution} of this population. The very existence of active galaxy
populations at high redshift of course implies that some fraction of
galaxies have had central black holes for a long time, 
but the detailed demographics are
less certain. In their recent work, \citet{Menou01} consider a
scenario in which the fraction of spheroids containing an MBH is a
function of redshift. In the following, such evolution is completely
degenerate with changes in the merger rate, but for definiteness we will
consider two scenarios, one in which the black hole mass function is
constant with time, and another in which the normalizing number density
follows a power law in $(1+z)$; this will allow us to combine a possible
evolution in the merger rate with evolution in MBH populations as a
single power law as in the previous subsection. In general 
we would certainly expect the
peak of the mass function to also evolve over cosmological timescales,
an issue we defer to later study.

\clearpage
\clearpage

\subsection{Black Hole Binary Dynamics}
\label{sec:BHmerg}

Once the merger of two galaxy halos occurs, we must then follow the
evolution of the central black holes. This is the subject of some
debate. The paucity of obvious close binary nuclei imply that mergers between
black-hole possessing-galaxies must be rare or short-lived. We know that the
host galaxies themselves have often experienced at least one ``major
merger'' in their lifetime. This is especially true of the most massive
ellipticals harboring the most massive black holes.  The recent study of
the dynamics of MBHs in merging galaxies given by \citet{Yu01}
builds on the seminal work of \citet{Begelman80}.

The evolution of the MBH binary proceeds in three stages. First, the
pair spiral into what becomes the common single core of the merger
remnant driven by dynamical friction, on a timescale given by
\begin{equation}
  \label{eq:tDF}
  t_{\rm DF} \simeq \left(\frac{4\times10^6\;\mbox{yr}}{\log N_*}\right)
  \left(\frac{\sigma_c}{200\;\mbox{km}\;\mbox{s}^{-1}}\right)
  \left(\frac{r_c}{100\;\mbox{pc}}\right)^2
  \left(\frac{M_\BH}{10^8 M_\odot}\right)^{-1}\;,
\end{equation}
where the core of the galaxy has radius $r_c$, velocity dispersion
$\sigma_c$ and contains $N_*$ stars, and $M_\BH$ is the mass of the
smaller black hole.

The binary continues to harden (inspiraling and gaining energy) first
through dynamical friction, subsequently by three-body interactions, and
finally as a hard binary, when the semimajor axis is given by
\begin{equation}
  \label{eq:hard}
  a\simeq \frac{GM_\BH}{4\sigma_c^2} = 2.8\; \mbox{pc}
  \left(\frac{\sigma_c}{200\;\mbox{km}\;\mbox{s}^{-1}}\right)^{-2}
  \left(\frac{M_\BH}{10^8 M_\odot}\right)\;,
\end{equation}
and the Keplerian period is given by
\begin{equation}
  \label{eq:Phard}
  P \simeq 4.4\times10^4\; \mbox{yr}\; \left(\frac{M_\BH}{10^8 M_\odot}\right)
  \left(\frac{\sigma_c}{200\;\mbox{km}\;\mbox{s}^{-1}}\right)^{-3}\; .
\end{equation}

At this separation, however, the timescale for evolution via
gravitational radiation, given by equation~(\ref{eq:tGW}), is much
longer than a Hubble time. So the typical pair must somehow evolve to a
much tighter orbit with $a\simeq0.02\;\mbox{pc}$ in order to {\em ever}
complete the merger. At this separation the Keplerian period is
$P\simeq30\;\mbox{yr}$ and orbital speed is $\sim$4000 km s$^{-1}$.  
Various dynamical mechanisms have been proposed
for this process.  \citet{Begelman80} first discussed the dynamics in
detail and noted the difficulty of reaching the rapid GW inspiral
regime, as the MBH binary empties the ``loss cone'' in the stellar
distribution function of stars with which the MBH binary can interact.
\citetalias{Rajagopal95} surveyed the field at the time, and decided
that while ``unassisted stellar dynamics'' would bring only a small
fraction to the GW regime, ``external perturbations may, however, cause
efficient inspiral.''  Examining these possible perturbations further,
\citet{Gould99} and \citet{ArmitageNatarajan02} revisited the earlier
idea \citep{Begelman80} that gas dynamics---the same physics as
planetary migrations---could be responsible for this evolution, and
indeed for fueling some or all quasar activity. Recently,
\citet{MilMer01} have done detailed N-body calculations 
of a MBH-MBH binary falling into a spherical potential of stars,
and discuss the issue of potential stalling of the binary coalescence.
\citet{Yu01} has examined in yet greater detail the stellar-dynamical
schemes in tri-axial galaxies. Yu concludes that indeed there should be
a sizable population of binary MBHs remaining in massive post-merger
galaxies, in the absence of such gas-dynamical effects, but notes the
difficulties of detecting the remnant binaries. 
On the other hand, \citet{ZhaoHaehnRees01} argue that the central
stellar distribution of galaxies is considerably more complicated (cuspy
and asymmetric) keeping the loss cone populated, possibly rescuing the
stellar-dynamic scenarios.

Another possibility is that the binary will be driven to the inspiral
regime by the presence of a third MBH from a second merger event. Tidal
forces from the
third body---or equivalent perturbations from a non-axisymmetric
galactic potential---can decrease the merger time by an order of
magnitude and significantly increase the
eccentricity of the binary via the ``Kozai mechanism''
\citep{blaesleesec02}. 

In the following, we will assume that some mechanism like gas physics is
indeed successful.  Thus, all relevant pairs enter the gravitational wave regime
and eventually shed their orbital energy in the form of gravitational
radiation. The final ten thousand seconds or so of this evolution are
thought to be the most energetic GW events in the (present-day)
Universe, and will be easily detected by the LISA satellite over a wide
range of MBH masses and redshifts \citep{HughesMN2001}. Here,
though, we are concerned with the quiescent quasi-Keplerian evolution
that precedes the infall events.

\subsection{Gravity Waves from Binary MBHs}
\label{sec:binaryGW}

Now, we will briefly review the emission of gravitational radiation from
binary black holes. First, we will need the amplitude of the
gravitational radiation emitted by a binary at distance $D$. This is
given by \citep{Peters63,Thorne}
\begin{equation}
  \label{eq:strain}
  h_{\rm rms}(r) = 4\sqrt{\frac{2}{5}} \left(\frac{G{\cal M}}{c^3}\right)^{5/3}
  \left(2\pi\over P_p\right)^{2/3} \left({c\over D}\right),
\end{equation}
where ${\cal M}=[M_1 M_2 (M_1 + M_2)^{-1/3}]^{3/5}$ is the ``chirp
mass'' of the system, and $P_p$ is the proper rest-frame period of the
binary. The observed frequency of the radiation in the $n$th harmonic
will be $f=nf_p/(1+z)$, where $f_p\equiv 1/P_p$.  
Only terms with $n\ge2$ are nonzero, and $n=2$
for the circular orbits to which we will restrict ourselves
hereafter \citep{quinlan96,quinlanhernquist97}. 
However, if the binaries are driven together by the presence of a third
MBH \citep{blaesleesec02}, the eccentricity may be significant and
populate higher-frequency harmonics. This would only serve to increase
the amplitude of gravitational radiation and populate the higher
harmonics (by an eccentricity-dependent factor). This amplification
would offset, or exceed, the loss resulting from expulsion of the
third body which is assumed to coalescence in our model.
Equation \ref{eq:strain} differs by a factor of $\sqrt{3/4}$ from that
given in \citetalias{Rajagopal95}. Instead of the ``characteristic
strain'' from \citet{Thorne}, which includes this factor to recover the
signal-to-noise for an earth-bound detector, we just use the
angle-averaged mean-square strain (\ie, eq.~[\ref{eq:strain}] is
$1/\sqrt{2}$ times the maximum rms strain). We point out for ease of
interpretation of this and later formulae that $(GM/c^3)$ has units of
time. In a cosmological setting, we replace $D$ with $c a_0 H_0
r(z)/H_0$.

We will also need the characteristic timescale of the emission, defined
in equation~(\ref{eq:taugwdef}). 
Using the Kepler formula, $f^2 a^3 = (2\pi)^{-2} G(M_1+M_2)$ and the
formula for energy loss from gravitational waves, the
gravitational wave timescale is \citep[\eg,][]{Peters63,ShapiroTeukolsky}
\begin{equation}
  \label{eq:taugw2}
\tau^{-1}_{\rm GW}\equiv  \left(\frac{1}{f_p}\frac{df_p}{dt_p}\right) = 
  \frac{96}{5}\left(\frac{G\cal M}{c^3}\right)^{5/3}
  \left[{\pi f(1+z)}\right]^{8/3}.
\end{equation}
where we have used $f_p=f(1+z)/n$ and set $n=2$ for circular orbits.

\section{Strain Spectrum}
\label{sec:strainspec}

Now we can combine all of the ingredients into an
observable strain power spectrum. We have
\begin{itemize}
\item $\nu(z)\; dz$, the volume merger rate observed at $z=0$ of 
 galaxies with MBHs located between $z$ and $z+dz$;
\item $\phi(M_\BH) \; dM_\BH$ the distribution of MBH masses, normalized
  to the number density of galaxies whose merger rate is given by $\nu(z)$;
\item $h_{\rm rms}(z, f, M_1, M_2)$, the strain observed at $z=0$ for a
  binary of given masses with frequency $f$ at redshift $z$; and
\item $\tau_{\rm GW}$, the gravitational radiation timescale.
\end{itemize}
Putting all this together using equation~(\ref{eq:Nzt}), the strain distribution is
\begin{equation}
  \label{eq:strain1}
  N(h,f,M_1,M_2) \; dh \; df\; dM_1 \; dM_2 = N(z, f, M_1, M_2)\;
  \frac{dz}{dh} dh \; df\; dM_1 \; dM_2
\end{equation}
where the distribution in redshift is

\begin{small}
\begin{eqnarray}
  \label{eq:strain1z}
  N(z,f,M_1,M_2) &=& \nu(z) \frac{\phi(M_1) \phi(M_2)}{\phi_\BH^2}
  (1+z)\tau_{GW}(M_1,M_2, z, f_p) \frac{1}{f} \nonumber\\
  &=& \frac{5}{96} \frac{\phi(M_1) \phi(M_2)}{\phi_\BH^2}
  \left(\frac{G{\cal M}}{c^3}\right)^{-5/3}
  \nu(z)(1+z)^{-5/3} \left({\pi f}\right)^{-8/3}
  \frac{1}{f}\nonumber\\
  &=& \frac{5}{96} \frac{4\pi c^3}{H_0^3} 
  \frac{\phi(M_1) \phi(M_2)}{\phi_\BH^2}
  \left(\frac{G{\cal M}}{c^3}\right)^{-5/3}
  R(z) \frac{\left[a_0 H_0 r(z)\right]^2}{E(z)(1+z)^{8/3}}
  \left({\pi f}\right)^{-8/3}  \frac{1}{f} \;.
\end{eqnarray}
\end{small}

First, note that this expression differs from equation~(20) of
\citetalias{Rajagopal95}; 
they neglected to include the factor of $(1+z)$ needed to convert
from proper rest-frame time interval $t_p$ to observed interval
$t$. Second, note that equation~(\ref{eq:strain1z}) factors into functions of mass, 
redshift and frequency. This means that marginal densities (integrals over one or
more of the variables) are easy to calculate and have the same
dependence on the remaining parameters as the unmarginalized
distribution. This will make the following calculations particularly
simple. However, it is certain that a fuller treatment of the model,
incorporating the likely dependence of the halo merger and MBH
coalescence rate simultaneously upon mass and redshift, including a
complete accounting for the probability of a given MBH binary of
successfully inspiraling to reach the GW regime, will no longer 
display this simplicity.

We now formulate the strain spectrum using the definitions of
equations~(\ref{eq:hc}--\ref{eq:h2N}):
\begin{equation}
  \label{eq:fullspec}
  h_c(f)^2 = f \int dh \;dM_1\; dM_2\;
  h_{\rm rms}^2(z,f,M_1,M_2) N(h,f,M_1,M_2)\; .
\end{equation}
We can simplify this expression considerably by using $N(h,\ldots)\; dh =
N(z,\ldots)\; dz$, which allows us to remove the ${dz}/{dh}$ factor in
equation~(\ref{eq:strain1}). Thus
\begin{equation}
  \label{eq:fullspecz1}
  h_c(f)^2 = f \int dz\;dM_1\;dM_2\;h^2_{\rm rms}(z,f,M_1,M_2) N(z,f,M_1,M_2).
\end{equation}
Using the results of equations~(\ref{eq:strain}--\ref{eq:fullspecz1}) and the
developments of the previous sections, the strain spectrum is given by
\begin{equation}
  \label{eq:fullspecz}
  h_c(f)^2 = \int dz \; \frac{4\pi c^3}{3}\left[\pi f(1+z)\right]^{-4/3}
  \langle \left(G{\cal M}/c^3\right)^{5/3}\rangle_{\BH} \frac{R(z)}{H_0 E(z)},
\end{equation}
where we use $\langle\cdots\rangle_\BH$ to denote averages over the MBH
population. In equation~(\ref{eq:fullspecz})
\begin{equation}
    \label{eq:Meff}
    \langle{\cal M}^{5/3}\rangle_{\BH} = \int\int dM_1 \; dM_2 
    \frac{\phi(M_1)\phi(M_2)}{\phi_\BH^2} {\cal M}^{5/3}
  = \left(2.3 \times 10^7 M_\odot\right)^{5/3},
\end{equation}
where we have used the MBH mass function of \S\ref{sec:BHpop},
and have assumed that the number density of MBHs and the halos whose merger rate is
described by equation~(\ref{eq:numrate}) are the same.

We can now also separate out the $z$ integration to see the simple form
of the result:

\begin{small}
\begin{eqnarray}
  \label{eq:fullsimple}
h_c(f)^2 &=&\frac{4\pi c^3}{3}\langle\left(G{\cal M}/c^3\right)^{5/3}\rangle_{\BH}\left(\pi f\right)^{-4/3}R_0
  \int \frac{R(z)}{R_0}\frac{dz}{H_0 E(z)(1+z)^{4/3}} \nonumber \\
& = & (1.8 h_0\times10^{-15})^2 
\langle\left(\frac{\cal M}{2.3\times10^7 M_\odot}\right)^{5/3}\rangle_{\BH}
\left(\frac{f}{\mbox{yr}^{-1}}\right)^{-4/3} 
\int \frac{R(z)}{R_0}\frac{dz}{E(z)(1+z)^{4/3}}\;.
\end{eqnarray}
\end{small}

In the last line we have set the number density of merging
galaxies and their present day merger rate to
\begin{eqnarray}
  \label{eq:nums}
  \phi_\BH &=& 1.0\times 10^{-3} h_{0}^3{\rm~Mpc}^{-3}\;, \qquad\mbox{and}
  \nonumber\\
  R_g &=& 0.09 {\rm~Gyr}^{-1}\;.
\end{eqnarray}
\citet{Phinney01}  derived an expression equivalent to 
equation~(\ref{eq:fullsimple}), and noted 
the weak dependence of $h_c(f)$ on
cosmology owing to a cancellation of $a_0H_0r(z)$ factors.
Only the $E(z)$ factor remains in the integral.  
However, the details of the distribution
$N(h,f)$ {\em do} depend more strongly on the cosmology.  

\begin{figure}[htbp]
\plotone{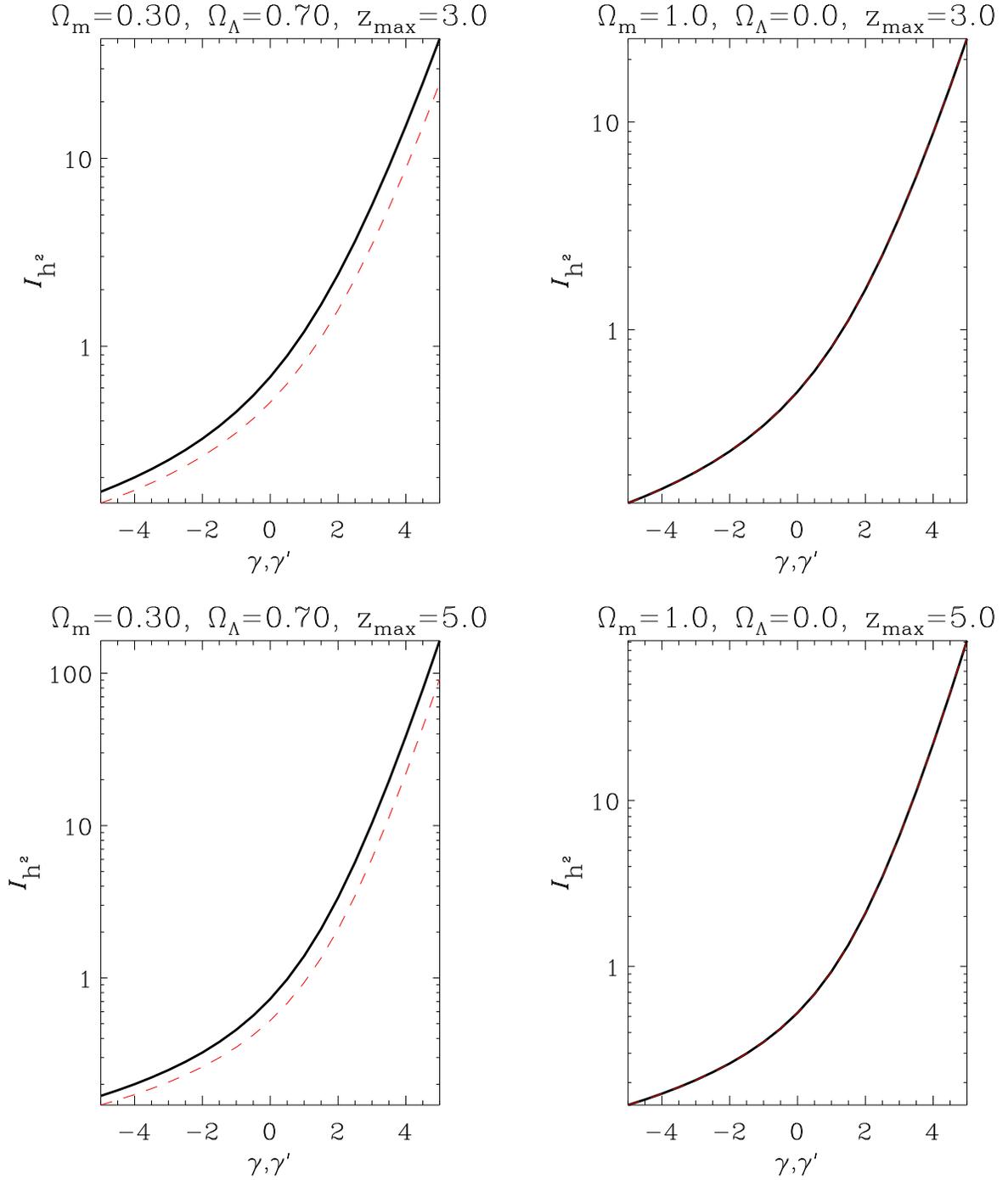}
  \caption{Redshift integral occurring in the MBH gravitational wave strain
    spectrum, equation~(\ref{eq:zint}), as in
    Figure~\ref{fig:totalrate}. The strain spectrum is proportional to
    this quantity.}
  \label{fig:strainint}
\end{figure}

In Figure~\ref{fig:strainint} we show the dimensionless integral,
\begin{equation}
  \label{eq:zint}
  I_{h^2}\equiv\int \frac{R(z)}{R_0}\frac{dz}{E(z)(1+z)^{4/3}},
\end{equation}
which is essentially the same as $\langle (1+z)^{-1/3}\rangle$ in
\citet{Phinney01}. We plot $I_{h^2}$ for a variety of cosmologies,
merger rates and black-hole mass-function histories. We parameterize the
latter two by the single exponent $\gamma$, defined such that
$\phi(M_1,z)\phi(M_2,z)R(z)\propto (1+z)^\gamma$ or $\propto
E(z)(1+z)^{\gamma'-3/2}$ as in \S\S\ref{sec:merger}--\ref{sec:BHpop}. 
The maximum redshifts of 3 and 5 used in Figure~\ref{fig:strainint}
are motivated by the distribution of objects with redshift which is discussed
below in \S\ref{sec:other}. 
For $\gamma\lesssim2$, we are most sensitive to $z<3$ and the cutoff is
irrelevant. For stronger merger-rate evolution, we are sensitive to more
distant events.
Figure~\ref{fig:strainint} shows the weak dependence upon both the
cosmology ($\Omega_m$ and $\Omega_\Lambda$) and the maximum redshift of
MBH binary formation.  The results also show the weak dependence on
whether we consider the merger rate as a power law in redshift (with
exponent $\gamma$) or time ($\gamma'$).  Similarly, if we allow this
exponent to vary by as much as one or two units, we again get very
little change.
The dependence on the maximum redshift is obviously somewhat stronger if we
increase the power-law exponent $\gamma$, thereby putting more events at
higher redshift.
There remains a
prefactor of $R_0$ in $h_c^2(f)$---the overall normalization of the
merger rate---which can have a strong effect. Of course, each of these
effects can change things by a factor of two.  In summary, the
uncertainty in $h_c^2(f)$ is at least an order of magnitude.

\clearpage
\clearpage

\subsection{The Strain Distribution}
\label{sec:straindist}

We would like to estimate the entire probability distribution of
$h_c^2(f)$ about this mean. The simplest way to approach this is to use the
method of \emph{characteristic functions}, which are simply the Fourier
transforms of probability densities. In order to do this, first consider
the quantity
\begin{equation}
  \label{eq:hc2}
  h_c^2(f) \frac{df}{f} = \int \; dz
  \; dM_1\; dM_2\; h^2_{\rm rms}(z,f,M_1,M_2) {\hat N}(z,f,M_1,M_2)  df\; .
\end{equation}
In this equation, the quantity ${\hat N}(z,f,M_1,M_2) \; dz \; df\;dM_1\;dM_2$ gives
the actual number in a cell of size $dz\times df\times dM_1\times dM_2$
at a given frequency and bandwidth $f$ and $df$. It can be shown (see
Appendix~\ref{appendix}) that the characteristic function of the
distribution of $h_c^2(f)\; df/f$ is given by
\begin{equation}
  \label{eq:charfun}
  \ln\varphi(t)=\int dz \; dM_1\; dM_2\; 
N(z,f,M_1,M_2)\left[e^{it h_{\rm rms}^2(z,f,M_1,M_2)}-1\right]\; df\;.
\end{equation}
While one can show that this integral exists for the cases
we are considering, it cannot be expressed in closed form. For
this reason, the distribution itself---the inverse Fourier transform
of $\phi(t)$---is quite difficult to calculate. 

To start, we can  use the fact that the coefficients of
the Taylor expansion of $\ln\phi(t)$ around $t=0$ give $\kappa_k$, the
`semi-invariant moments' or `cumulants' of the distribution (\ie, the
mean, variance, skewness, kurtosis, etc. for $k=1,2,3,4,\ldots$), which
we can read off as
\begin{equation}
  \label{eq:kappak}
  \kappa_k(f) = \int dz \;dM_1 \; dM_2 \left[h_{\rm rms}^2(z,f,M_1,M_2)\right]^k N(z,f,M_1,M_2) \; df\;.
\end{equation}
Note that these are the moments of $h_c^2(f)\; df/f$, not $h_c^2(f)$
itself, which can only be measured with a finite bandwidth. 
For example, the variance of $h_c^2(f){df}$ is
\begin{equation}
  \label{eq:varh2fdf}
  {\rm var}\left[h_c^2(f) \; df \right] = 
  f^2 \; {\rm var}\left[h_c^2(f) \; df/f \right] =
 f^2 \int dz \; dM_1\; dM_2\; h_{\rm rms}^4 N(z,f,M_1,M_2) \; df\;.
\end{equation}
We measure $h_c^2(f)$ itself with some finite bandwidth,
$\Delta f$, as $h_c^2(f)\simeq\int_{\Delta f} h_c^2(f)\; df/{\Delta
  f}$. The variance on this quantity decreases as the bandwidth increases.

The variance of the MBH density $N(h,f)$ under consideration is
\begin{equation}
  \label{eq:varh2}
  {\rm var}\left[h_c^2(f)\right] = 4\pi c^3 H_0 \frac{32}{5} 
  \left(\frac{\Delta f}{f}\right)^{-1}
  \langle\left(G{\cal M}/c^3\right)^5\rangle_{\BH}
  \int dz\;\left[a_0 H_0 r(z)\right]^{-2} R(z)/E(z)\;.
\end{equation}
As before, this separates into an average over the MBH mass function and
an integral over redshift. Considering the latter, we see that the
integral in question diverges as $z\to0$, with the integrand behaving as
$dz/z^2$. Indeed, all of the integrals with $k>1$ diverge: the moments
do not exist. Nonetheless, the characteristic function and the
distribution itself remain well-defined.  This divergence occurs because
the strain from a single event is proportional to the inverse of the
distance; although the mean-square accumulated strain (the power
spectrum) converges, the dispersion from
small distances diverges.

We have performed numerical experiments with the simpler problem of a
uniform MBH binary distribution in a non-expanding, Euclidean universe,
so $N(z)\propto z^2$ and $h_{\rm rms}^2(z)\propto1/z^2$ for all
$z<z_{\rm max}$---an
approximation to the situation at $z\ll1$. We find that
the distribution of strains is highly skewed, with a power-law
tail toward larger amplitudes. 
This tail is due to the possible contribution of low-redshift
pairs which give a contribution $h^2(z)\propto1/z^2$.

\subsection{Other Quantities}
\label{sec:other}

\begin{figure}[htbp]
\plotone{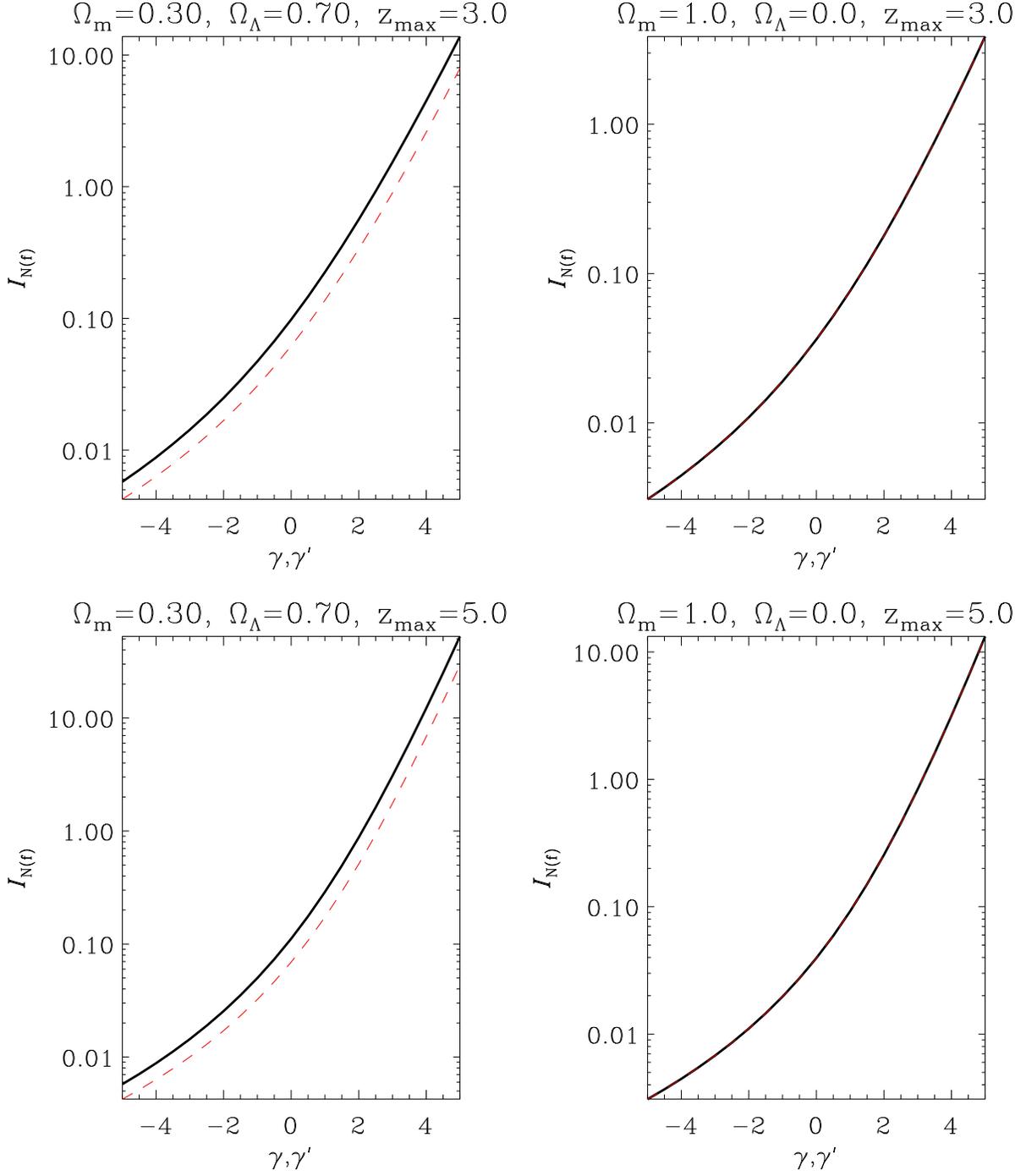}
  \caption{The dimensionless integral occurring in $N(f)$,
    equation~(\ref{eq:Nf2}), for a variety of scenarios as in
    Figure~\ref{fig:totalrate}.}
  \label{fig:Nf}
\end{figure}

Our derivation also allows us to calculate other quantities related to
the distribution of strains. First, we can ask what is the number of
binaries in a frequency interval,
\begin{equation}
  \label{eq:Nf}
  N(f)\; df = \int dz\;dM_1\;dM_2 \; N(z,f,M_1,M_2)\; df \; .
\end{equation}
Using equation~(\ref{eq:strain1z}), and again under the assumptions we have
made so far that the $\phi_\BH$ is independent of redshift, the
$z$ integration separates out, and we see first that $N(f)\; df/f \propto
f^{-8/3} df/f$, a steeply-falling function of frequency. In full,
\begin{equation}
  \label{eq:Nf2}
  N(f)\; df =  5.6\times10^5\; df/f\; 
    \langle\left(\frac{\cal M}{4.1\times10^6 M_\odot}\right)^{-5/3}\rangle_{\BH}
    \left(\frac{f}{\mbox{yr}^{-1}}\right)^{-8/3} 
    \int dz\; \frac{R(z)}{R_0}\frac{[a_0 H_0 r(z)]^2}{E(z)(1+z)^{8/3}}\;. 
\end{equation}
$\langle{\cal M}^{-5/3}\rangle_{\BH}$ is $\left(4.1 \times 10^6
  M_\odot\right)^{-5/3}$ for our MBH mass function which is different
from the value above owing to a different weighting function.  In
Figure~\ref{fig:Nf} we show the dimensionless integral occurring in
this expression for a variety of cosmologies and merger histories. The
stochastic GW background at nHz (yr$^{-1}$) frequencies is therefore the
result of nearly a million simultaneous binaries across the Universe.

\clearpage
\clearpage

\begin{figure}[htbp]
\plotone{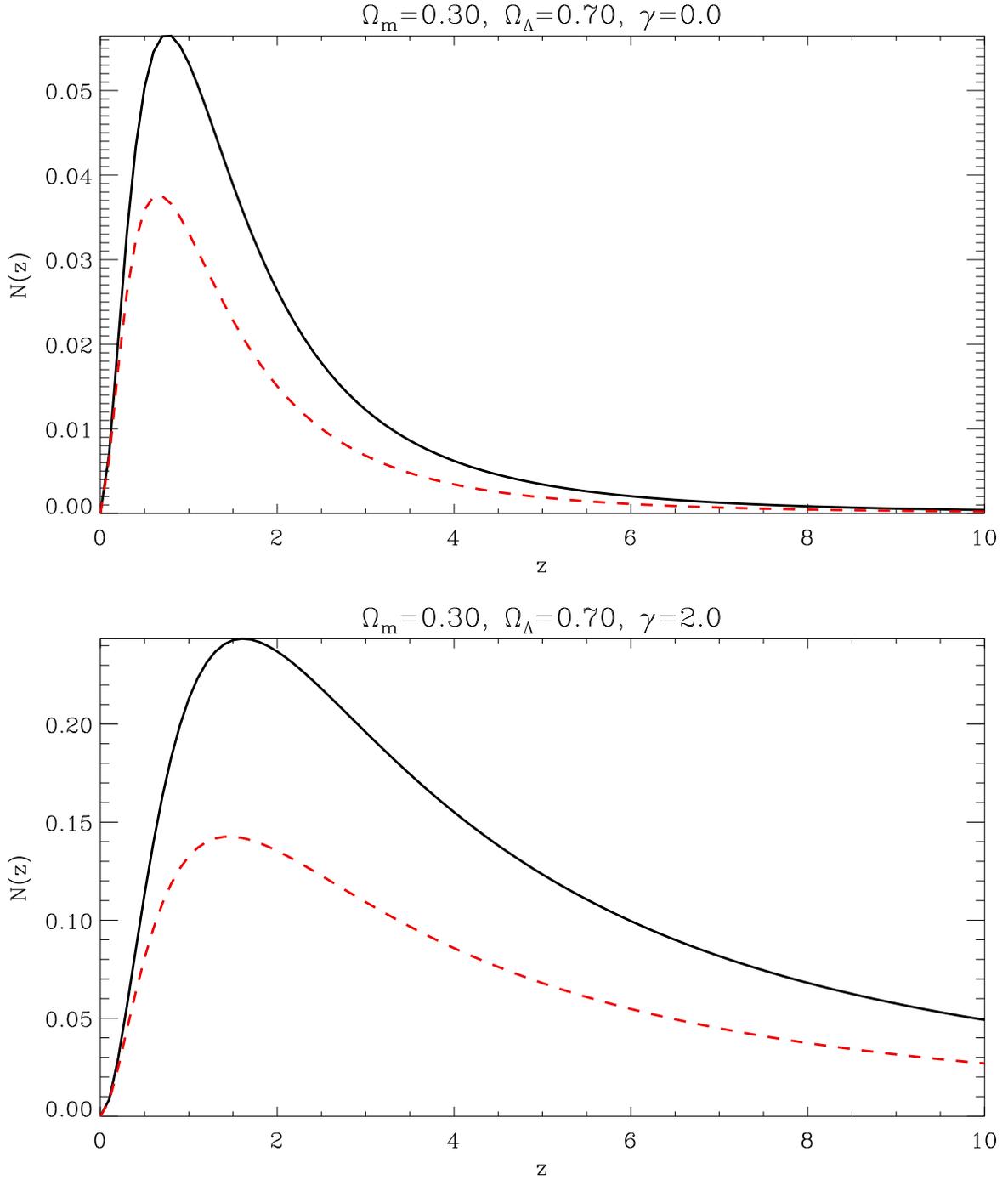}
  \caption{The distribution of redshifts probed by MBH Binaries,
    equation~(\ref{eq:Nz}), for two sets of cosmological parameters and
    two choices of parameterizing galaxy merger rate. The solid lines
    parameterize the merger rate as a power-law in redshift; the dashed
    lines as a power law in time, as in Figure~\ref{fig:totalrate}.}
  \label{fig:zdist}
\end{figure}

\begin{figure}[htbp]
\plotone{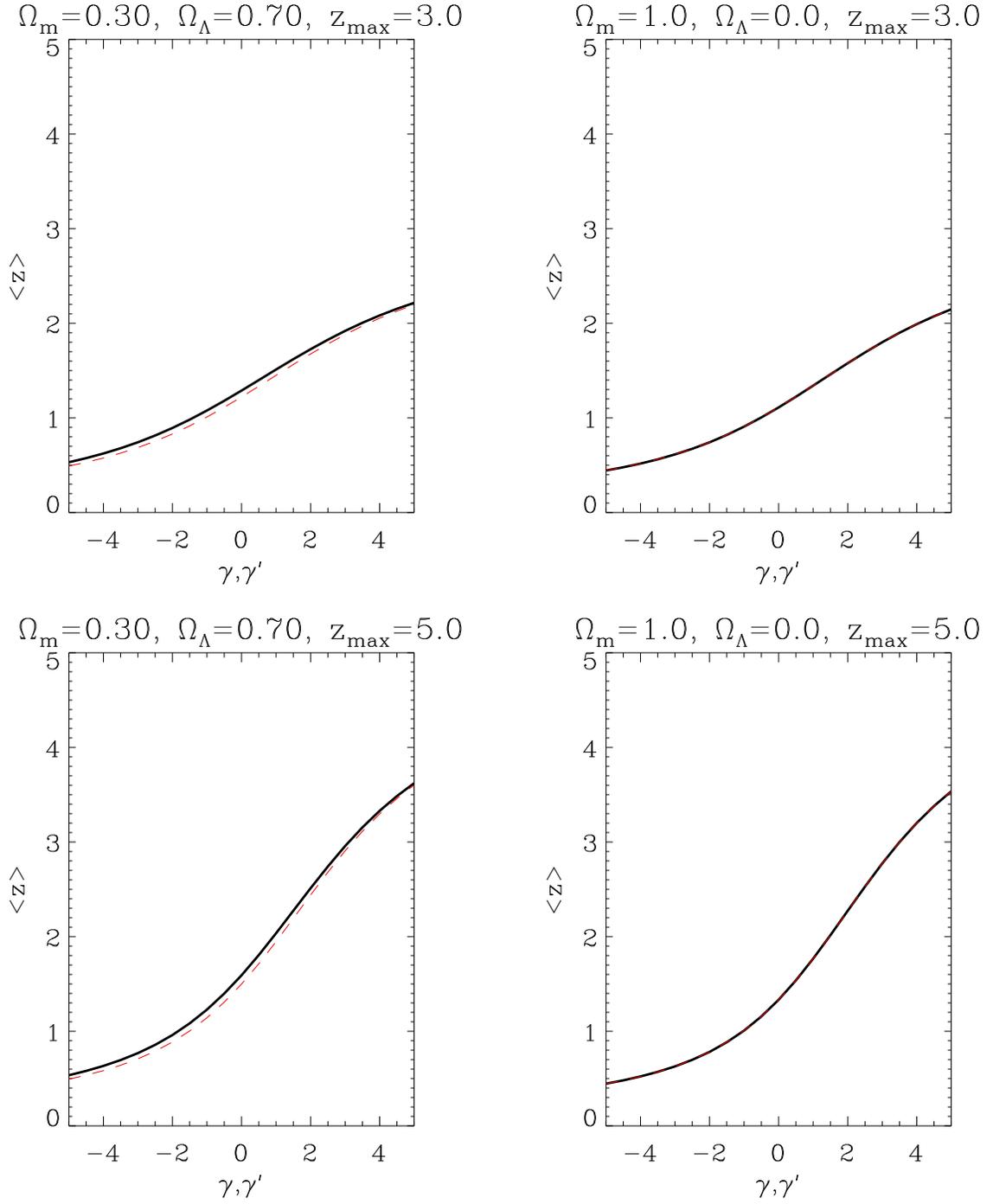}
  \caption{The average redshift from which the MBH gravitational wave signal
    comes, equation~(\ref{eq:zf2}), for a variety of scenarios as in
    Figure~\ref{fig:totalrate}.}
  \label{fig:zav}
\end{figure}

Similarly, we can ask, what is the range in 
redshift that we are probing? This is just 
\begin{equation}
  \label{eq:Nz}
  N(z) \propto \frac{R_g(z)}{R_0}\frac{[a_0 H_0 r(z)]^2}{E(z)(1+z)^{8/3}}.
\end{equation}
which we show in Figure~\ref{fig:zdist}. 
For larger values of the exponent ($\gamma\gtrsim2$), relatively more
events occur at higher redshift, and the dependence of the observed
strain on the maximum redshift will be stronger.
More specifically, we can calculate the mean redshift,
\begin{equation}
  \label{eq:zf}
  \langle z(f) \rangle = \frac{\int dz \; z\; N(z,f)}{\int dz \; N(z,f)}\; .
\end{equation}
Again, the integrals separate and we find that this average redshift is
independent of frequency:
\begin{equation}
  \label{eq:zf2}
  \langle z \rangle = \frac
  {\int dz\; R(z) \left[(a_0H_0r)^2/E(z)\right](1+z)^{-5/3}}
  {\int dz\; R(z) \left[(a_0H_0r)^2/E(z)\right](1+z)^{-8/3}}-1\;.
\end{equation}
This is independent of the details of the MBH population and the
normalization of the merger rate, but dependent on the merger rate
evolution and the cosmology.
In Figure~\ref{fig:zav} we show the average redshift for a variety of
cosmologies and merger histories.

\clearpage

\clearpage

\clearpage

In the previous sections we calculated the strain spectrum, $h_c^2(f)$,
and variants. Using the above developments, we can calculate a slightly
different quantity: the average strain amplitude that
contributes to $h_c^2(f)$. This is just 
\begin{eqnarray}
  \label{eq:avgstrain2}
  \langle h^2(f) \rangle &=& \frac{h_c^2(f)}{N(f)}
  \propto f^{4/3},
\end{eqnarray}
which rises owing to the more rapid decrease of number density
with frequency relative to the spectrum amplitude decrease.
That is, higher frequency signals are from small numbers of rare,
powerful events in comparison with lower frequency signals.

Finally, we can also calculate quantities related to the mass
distribution that contribute to the strain spectrum. Again due to the
factorization of $N(M_1, M_2, z, f)$ these are independent of redshift
and frequency. For any function of the masses, we see that
\begin{eqnarray}
  \label{eq:fmassavg}
  \langle F(M_1, M_2) \rangle = 
  \frac{\langle F(M_1, M_2) {\cal M}^{-5/3}\rangle_\BH}
  {\langle {\cal M}^{-5/3}\rangle_\BH}\;,
\end{eqnarray}
where as above $\langle\cdots\rangle_\BH$ refers to averages over the
MBH distribution. One particular quantity of interest is simply the
average chirp mass:
\begin{equation}
  \label{eq:avgchirp}
  \langle {\cal M} \rangle = 2.9\times 10^6 M_\odot.
\end{equation}
which, as would be expected, is close to the average of the MBH
distribution itself (Fig.~\ref{fig:BHmassfn}). Another is the mass ratio,
\begin{equation}
  \label{eq:avgmassratio}
  \langle q \rangle \equiv \langle M_1/M_2 \rangle = 11\;.
\end{equation}
Because the MBH mass distribution is itself so wide
(Fig.~\ref{fig:BHmassfn}), the gravitational waves sample a fairly wide range
of black-hole mass ratios. In particular, we are sensitive to minor
mergers as well as major mergers. However, these quantities change to
$\langle{\cal M}\rangle=5\times10^6M_\odot$ and $\langle q\rangle=1.7$
if a minimum black hole mass of $10^6 M_\odot$ is assumed. Such a
minimum hole mass is warranted based on considerations of the strong
dependence of dynamical friction time scale on mass \citep{XuOstriker94}.
The values in equations~(\ref{eq:avgchirp}--\ref{eq:avgmassratio}) can
be expressed as individual masses $\langle M_1,M_2\rangle\sim (1.3\times
10^7,1.1\times 10^6)~M_\odot$, or $(7.4\times 10^6, 4.4\times
10^6)M_\odot$ if the distribution is cut off at $10^6M_\odot$.

In a similar vein, we can weight these functions by the signal amplitude
they produce, rather than the number of binaries contributing. Since
$h_{\rm rms}^2\propto {\cal M}^{10/3}$ (eq.~[\ref{eq:strain}]), this
changes the exponents in equation~\ref{eq:fmassavg} from $-5/3$ to
$+5/3$. Such a weighting gives a signal-weighted average chirp mass of
$9\times10^7M_\odot$ and mass ratio of $6$. Again, we see that a
small number of high-mass black holes are responsible for the bulk of
the signal. Similarly, the signal-weighted average redshift is somewhat
lower than the number-weighted redshift of Figure~\ref{fig:zav}.

\subsection{Results}
\label{sec:specresults}

\clearpage
\begin{figure}[htbp]
\plotone{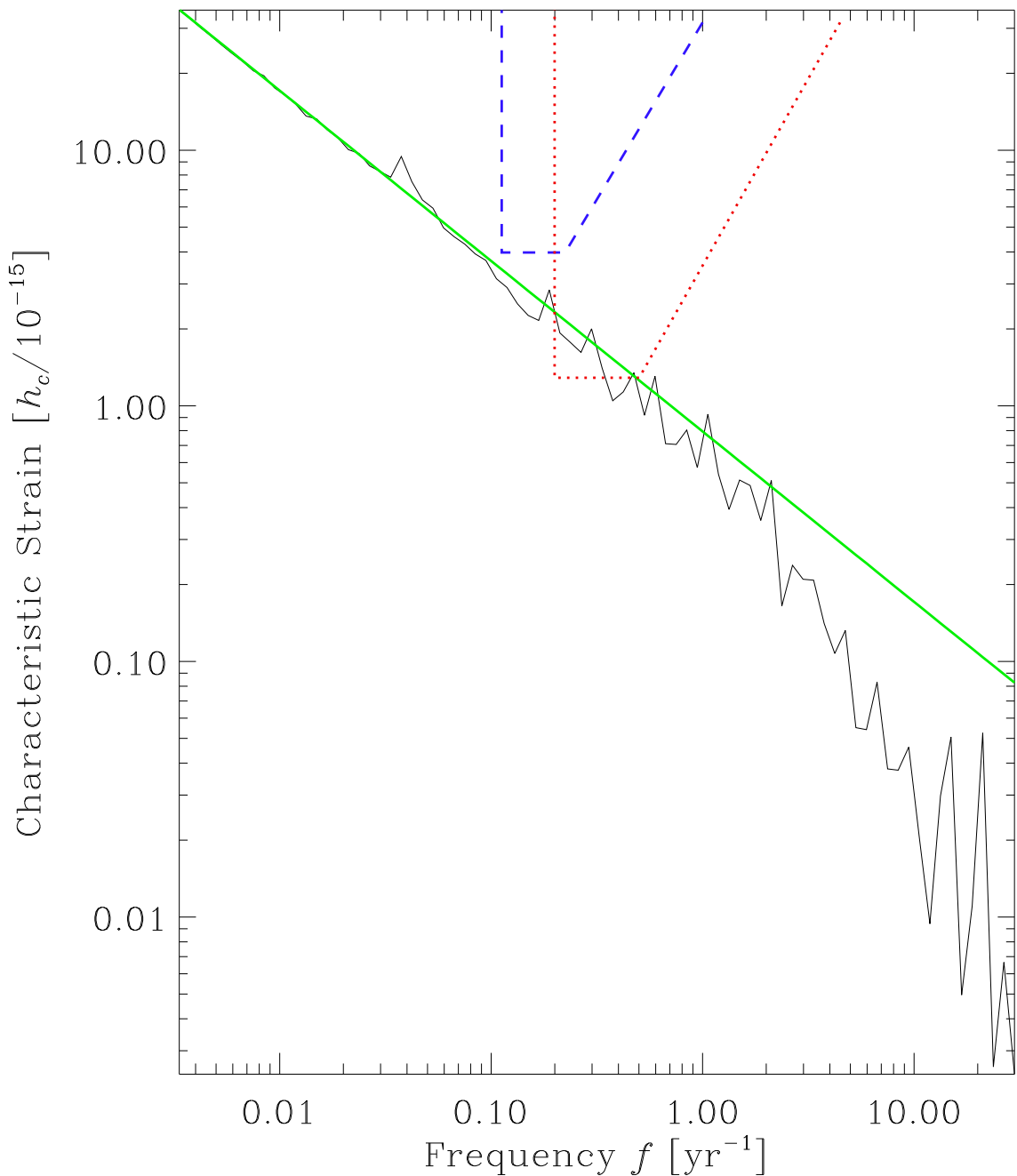}
  \caption{Characteristic strain spectrum $h_c(f)$ for the fiducial
    models discussed in the text, along with Monte-Carlo realizations.
    The upper thick curve and the associated realizations have
    $\gamma=2$; the lower set of curves have $\gamma=0$. The blue dashed
    line gives the current best limits on the gravitational wave
    background from pulsar timing observations. The dotted line
    shows the expected limits from a Pulsar Timing Array, after
    operation for $\sim8$ years.}
  \label{fig:strainspec}
\end{figure}
\clearpage

In Figure~\ref{fig:strainspec} we consider two models
of galaxy mergers and the black hole population, and show the resultant
spectrum of characteristic strain along with several realizations of a
Monte-Carlo simulation of the spectrum. 
We perform the Monte Carlo by simply Poisson-sampling from $N(z, f, M_1,
M_2)$ on a grid of the parameters and summing the contributions of
$h_{\rm rms}^2(z, f, M_1, M_2)$.  We have chosen $\Omega_m=0.3$,
$\Omega_\Lambda=0.7$, the constant halo merger rate given in
equation~(\ref{eq:numrate}), the MBH mass function from
equation~(\ref{eq:BHMF}), and a maximum redshift of $z_{\rm max} = 3$.
The upper set of curves have a relatively strong power-law index
$\gamma=2$ describing the galaxy merger rate; the lower curves have
$\gamma=0$.  We have assumed a bandwidth of $\delta f/f\simeq0.12$.  We
note that at higher frequencies the simulation is somewhat below the
expected mean spectrum. This is due to the high-amplitude tails
mentioned in the discussion of the full strain distribution,
\S\ref{sec:straindist} above: small number statistics and numerical
inaccuracies mean that we miss the large spikes.

We expect the actual dispersion to be greater at higher frequencies---smaller numbers of
objects are contributing more due to the steeply-falling function $N(f)$
(eq.~[\ref{eq:Nf}]). However, just as the mean of our simulation falls
below the expectation value, so does the dispersion around the mean: we
miss the large excursions. This is a larger effect for the
relatively high value $\gamma=2$ than at $\gamma=0$, representing
comparatively fewer merger events at low redshift. 

However, a central result of this paper is that the \emph{slope} of the
spectrum is independent of all of this. As in \citet{Phinney01}, we find
that $h_c^2(f)\propto f^{-4/3}$, or $S_h(f)\propto f^{-7/3}$, or
$\Omega_{GW}(f)\propto f^{2/3}$.  The overall amplitude, however, will
vary up and down by the factors calculated above and shown in
Figure~\ref{fig:strainint}. As mentioned above, this means that, on the
one hand, we cannot use the information from any observed slope to limit
the physics. On the other hand, it potentially provides a check that we
are observing the background from a coalescing population (although it
is not sensitive to the details of that population). Indeed, other
proposed sources of stochastic gravitational waves often have different
slopes. Early universe cosmological backgrounds, for example,
generically have $\Omega(f)\propto\textrm{const}$ and hence
$h_c^2(f)\propto f^{-2}$.

\section{Gravity Wave Detection with a Pulsar Timing Array}
\label{sec:PT}

We turn now to the experimental approaches to detection of the stochastic
background from MBH binaries. We begin with a statement of the response
of pulsar timing to a plane GW. In the following subsection we discuss
the angular and temporal basis functions required
for detection of a stochastic background. In the final subsection
the current limits and future prospects are discussed in relation
to the expected signal level presented in Figure~\ref{fig:strainspec}.

\subsection{Plane Wave Response}
Precision timing of the arrival of pulses from a pulsar
allows the direct detection of gravitational radiation traversing the sightline
\citep{Sazhin78,Detweiler79}. The spacetime metric strain perturbation
$h$ can be viewed as a perturbation from unity of the index of refraction 
for electromagnetic radiation in flat space. For
a plane GW traversing the entire sightline to a pulsar the
pulse arrival time is advanced, or retarded, in proportion to the
incomplete cycles of the wave traversed by the pulse at the pulsar
and in the solar system. Observable effects are possible only when
the GW amplitude, or state of the spacetime metric, changes locally or
at the pulsar or both in ways distinguishable from pulsar spin and other
parameters. A useful analytic approach is to formulate
the apparent redshift $Z$ of a pulsar spin frequency
which arise from gravitational radiation. Observationally $Z$ is estimated 
by taking the derivative of the residuals that result
after fitting pulse arrival times to a model for the pulsar's
spin properties and astrometric coordinates, and binary parameters if relevant. 
For a single plane wave the observable may be written as
\begin{equation}
Z(\alpha,\beta,\gamma;t_{\rm r}) 
  =  {(\alpha^2-\beta^2)\over 2(1+\gamma)}
     [h_+(ct_{\rm r}) - h_+(ct_{\rm e}-\gamma l)] +
     {\alpha\beta\over (1+\gamma)}
     [h_\times(ct_{\rm r}) - h_\times(ct_{\rm e}-\gamma l)], 
 \label{eq:Z}
\end{equation}
where $(\alpha,\beta,\gamma)$ are the direction cosines of the pulsar
with respect to the GW, $(h_+,h_\times)$ are the polarization components of the
GW, and $(t_{\rm e},t_{\rm r})$ are the times of emission (e) and
reception (r) of the pulses from the pulsar which is at a distance
$l$. 

The $t_{\rm r}$ terms in equation~(\ref{eq:Z}) 
will be correlated between different pulsars. Therefore timing measurements
of an array of pulsars across the sky, the Pulsar Timing Array, creates 
a gravitational wave telescope that is sensitive to a spectrum of waves with 
periods less than the duration of the measurements
which is years, or $\sim$ 10 nHz \citep[\eg][]{Backer93}.
The five degrees of freedom in equation~(\ref{eq:Z}) correspond to the five
degrees of freedom in the trace-free space-time metric. 
The $t_{\rm e}$ terms in equation~(\ref{eq:Z})
will be uncorrelated between different pulsars,
and yet will have comparable amplitude. These terms create an irreducible
noise background in addition to any limitations from the measurement errors
and properties of the pulsars themselves or the intervening turbulent plasma 
through which the signals propagate. \citet{Lommen01} apply these ideas to
a search for plane waves from both our Galactic Center and nearby galaxies
which host MBHs based on the hypothesis that the central objects are binary.

\subsection{Fitting Algorithms for a Stochastic Background}

In the preceding sections we considered the production of a
stochastic background of gravitational radiation from coalescences of
MBH binaries. This is summarized in the form of a characteristic strain spectrum, $h_c(f)$,
Figure~\ref{fig:strainspec}.
For the purposes of analysis of timing data from a spatial array
of pulsars we need to consider the full tensor perturbation field 
$h_{ij}({\bf x},t)$.  All elements are statistically independent owing
to the random superposition of many sources from many directions.
This superposition destroys the simple angular plane wave pattern described
in equation~(\ref{eq:Z}).
However, \citet{Burke75} has shown that 5/8 of 
the variance in the stochastic fluctuations can be extracted using
the 5 quadrupole spherical harmonic $Y^l_m$ functions as the angular
basis. 

Temporal modulations of each spherical harmonic term can be described using 
either Fourier frequency coefficients as the basis, 
or polynomials \citep{Foster90},
or orthogonal polynomials \citep{Stinebring}. The polynomial approaches are
particularly suited to this analysis owing to both
the steep spectrum expected for the MBH-MBH GW 
emission (eq.~[\ref{eq:fullsimple}]) and the need to fit for the spin
parameters of the stars. 
Given any choice of the temporal basis function, the amplitude 
$h_{\rm rms}(f)$ for any temporal term is formed by the square root of the
sum of the squares of the 5 spherical harmonic coefficients. Conversion
to $h_c(f)$ requires consideration of the spectral window function
which will be addressed in a later work.

A second approach to the angular basis has been suggested by \citet{Hellings90}. 
The tensor field stated above can be transformed
into a spectrum of waves $\tilde h_{ij}({\bf q},f)$. 
These can be decomposed into
waves traveling along the three cardinal directions, each with
independent polarizations:
\begin{equation}
\left[\begin{array}{ccc}
h_{11}  &h_{12} &h_{13} \\
h_{12}  &h_{22} &h_{23} \\
h_{13}  &h_{23} &h_{33}
\end{array}\right]=
\left[\begin{array}{ccc}
h_{x,+}       &h_{x,\times} &0 \\
h_{x,\times}  &-h_{x,+}      &0 \\
0             &0            &0
\end{array}\right]+
\left[\begin{array}{ccc}
h_{y,+}       &0            &h_{y,\times}  \\
0             &0            &0             \\
h_{y,\times}  &0            &-h_{y,+} 
\end{array}\right]+
\left[\begin{array}{ccc}
0             &0            &0             \\
0             &h_{z,+}      &h_{z,\times}  \\
0             &h_{z,\times} &-h_{z,+} 
\end{array}\right]
\end{equation}
In this case one of the six terms in the decomposed format is not
independent owing to the trace-free property. The Doppler patterns
of the three waves can then be summed to create Doppler patterns
for each tensor element.
Thus the full variance of the stochastic background can be sampled.
As in the spherical harmonic case each term will be
temporally modulated, and  the
quantity $h_{\rm rms}$ is formed by the square root of the sum of the
squares of 5 independent terms.  

\subsection{Current Limits and Future Prospects}

While the preceding subsection outlines approaches to detection of
the GW background using an array of pulsars, measurements from a single
object can establish an {\it upper limit} on the background.
\citet{Kaspi94} has stated the best limit 
at nHz frequencies using precision timing
of PSR B1855+09 relative to the world's best atomic time scale. A second
pulsar in their study, PSR B1937+21, was shown to be unstable on long time
scales, an effect they attribute to internal structure of that neutron
star. \citet{Thorsett96} improved the limit using further statistical 
considerations. At a typical frequency of 5 nHz (8~yr period) they conclude
that the energy density in gravitational radiation per logarithmic frequency
interval is less than $\Omega_gh^2\sim 6\times 10^{-8}$, where $h$ in this expression is
the ratio of Hubble's constant to 100 km s$^{-1}$.

Energy density
scales quadratically with the time derivative of the metric strain.
Therefore the level of $\Omega_gh^2$ that can be detected by pulsar
timing scales quadratically with the rms timing residual $R$, and as $T^{-5}$
with the duration of the experiment $T$, which is the
inverse of the minimum frequency sampled.
Recently \citet{LommenBackerAAS} have extended the Kaspi study of PSR B1855+09
by more than doubling the experiment duration to 17 y. 
The new upper limit on energy density is roughly
an order of magnitude lower at frequencies near 5 nHz. A full report on
this work is in preparation. An approximate statement of the Kaspi-Lommen
limit on characteristic strain 
is given in Figure~\ref{fig:strainspec}, which shows that the
measurements are approaching a level of significance given our current
model of the Universe. An alternate way of stating this is that the measurements
are placing useful constraints on the uncertain parameters in our model.
The ratio of the pulsar detection level to the characteristic strain 
spectrum level scales as $T^{13/6}$.
 
The Pulsar Timing Array experiment, which will use both existing and new
data sets, can improve on the \citet{Kaspi94} result in three obvious ways:
(a) smaller timing residuals owing to combined data sets,
new and upgraded telescopes and better
data acquisition techniques; (b) more objects which both
provide the capability
of actual detection as opposed to upper limits and, if sufficiently
numerous, provide a ``root N'' advantage; and (c) longer experiment
duration. In Figure~\ref{fig:strainspec} we show a reasonable goal for the future:
200 ns timing precision over eight years. Current measurement series
are already achieving this level of precision for a few objects.

\section{Summary}

We have calculated the spectrum of the stochastic background of gravitational
radiation from the Universe of coalescing binary black holes in the
centers of galaxies with simple 
parameterizations of the current uncertainties. This is followed by a
discussion of the influence of the stochastic background
on precision timing measurements of pulsars which includes upper limits
on the background and use of an array of pulsars for direct
detection. This work was motivated by improvements both in our knowledge
of the present-day massive black hole population and the rate of galaxy
mergers as well as new pulsar measurements.

Following the approach of past authors, we construct
the spectrum from coalescing binary MBHs from: the galaxy merger
rate, the black hole population demographics amongst galaxies, 
MBH binary dynamics and MBH binary gravitational-wave emission. 
Our galaxy merger rate 
is based on observations of close pairs and an estimate of their
dynamical friction time scale. We convert from the galaxy mass function
to the black hole mass function using recent determinations of
the correlation between black hole mass and spheroid mass.

Our ``fiducial'' model has rapid evolution in the merger rate per unit
time per galaxy as a function of redshift, and a constant mass function
of MBHs out to $z=3$.  The characteristic strain spectrum predicted for
this model, $h_c(f)\sim10^{-16} (f/{\rm yr}^{-1})^{-2/3}$, is just below
the latest observational limits at $f\sim0.2{\rm~yr}^{-1}$. Whereas the
slope and general character of this prediction agree with the earlier
work of \citet{Rajagopal95}, our predicted amplitude is somewhat higher.
This increase is the result of the order of magnitude increase in the
local MBH number density, which is tempered by a somewhat lower merger
rate at high redshift.

With our formulation we can calculate other quantities. The number of
binaries contributing for unit bandwidth ($\delta f/f\sim1$) at nHz
frequencies is approximately $10^6$. It is evident from our simulations
that the variance in the strain spectrum in the same frequency band is
roughly 50\%, considerably larger than the $10^{-3}$ that would be
expected if all events were weighted equally. This indicates that the
spread around the mean is due to the possibility of a small number of
high-amplitude, nearby events.  The median redshift for the binaries
contributing to the spectrum is $\le 1$ if the merger rate evolves
relatively slowly with redshift ($\gamma\lesssim2$), but can be large
for strong evolution of the merger rate. The mean ``chirp'' mass is
$3\times 10^6 M_\odot$ and the mean mass ratio is $11$; the individual
masses are then $10^7$ and $10^6~M_\odot$.

Our calculations allow us to see directly how the
remaining lack of understanding of some of these physical processes
impacts our predictions. These are areas for future research. 
The most important factors are:
\begin{itemize}
\item The value of the present day galaxy merger rate is a matter of
  considerable debate. The value we have quoted, 
  $0.09\; {\rm Gyr}^{-1}$ per galaxy from \citet{Carlberg00} is known to
  no better than 50\%. Moreover, this value is measured for massive
  $L_*$ galaxies; does it apply to the wider range harboring MBHs?
\item The evolution of the merger rate for moderate redshifts
  ($z\lesssim1$) is even less well-known.
  Looking back still further, the evolution of the merger rate at
  $z>1$ is yet more difficult to pin down. The high redshift surveys,
  in progress and planned, are crucial for progress.
\item We have not taken into account the details of MBH binaries
  dynamics: under what circumstances, if any, do MBH binaries make it to
  the GW regime? We have assumed that all galaxy mergers promptly lead
  to coalescence of their central black holes. A more complex model
  needs to consider both a delay between merger and coalescence which
  may depend on mass and the possibility of expulsion in a triple system.
\item Throughout, we have simplified the MBH coalescence rate by
  splitting the MBH population from the galaxy merger rate, as in
  equation~(\ref{eq:nuz}). Although this ansatz is useful for examining
  the dependences of the results upon the various ingredients, a full
  treatment of the time- and mass-dependent merger rate of galaxies (or
  the black holes directly) needs to be included. Appropriate results
  could come from considerably improved observations, N-body
  simulations, or formalisms such as the Press-Schechter calculation of
  the halo mass function.
\item Finally, we have assumed a simple model for the black hole population,
  combining the spheroid mass function and MBH demographics, as in
  \S\ref{sec:BHpop}. This does not take into account more recent
  results relating the MBH mass with the gravitational potential of its
  parent spheroid via the velocity dispersion
  \citep{gebhardt,ferrarese}. Moreover, we assume that the evolution in
  the MBH mass function can be modeled simply as a scaling of the number
  density, rather than the perhaps more realistic scenario where the
  mean and spread of the mass also evolves.
\end{itemize}

In future work, we plan on addressing these issues.  
Many of the observational uncertainties in the model are at high
redshift.  Modern `semi-analytic' models
\citep[\eg,][]{Cattaneo99,HaehneltKauffmann00} go a long way toward
elucidating these issues within the context of particular scenarios for
structure formation. Observationally, the quasar luminosity function
\citep{Boyle2QZLF00,FanSDSSQSOLF01}, OH Megamasers \citep{Darling02} and
``X-type'' radio sources \citep{MerrittEkers}
are other tools for understanding
the high-redshift MBH population. For this work, a detailed
high-resolution search for more nearby MBH binaries themselves, perhaps
in the form of double-nucleus QSOs \citep[but see][]{KochBinQSO99} and
Ultra-Luminous Infrared Galaxies \citep{MurphyULIRG01}, will be a
crucial
tool.  Analytical and N-body studies of the dynamics and energetics of
MBH binaries within galaxies will provide population information on
MBH-MBH mass ratios and orbit sizes crucial to future work.  These will
allow us to also constrain the dynamics of the MBH binary population:
if and when do they stall prior to the BW regime?

We can, of course, use our current model to at least see the effect of
these uncertainties. We can find a lower limit to the effect of MBH
mass-function evolution by using a suitable initial mass function for
the whole redshift range. If the average MBH mass has increased by an order of
magnitude (say), then this would lower the characteristic strain
(eq.~[\ref{eq:fullsimple}]) by no more than $10^{5/6}\approx7$.

The current experimental limit on the low-frequency gravitational wave
background has improved over the pioneering work of \citet{Kaspi94}, and
will be accurately stated by Lommen \& Backer (in preparation). The
new limit is lower at lower frequencies while the expected spectrum 
is rising. This new result
will constrain parameters of our model spectrum on the high side of
the current estimate. Ongoing work with an array of millisecond pulsars has
the prospect of significant improvement of detection capability in the
coming decade. Techniques for optimal fitting of Pulsar Timing Array
data need to be further developed to meet the demands of the
new measurements.

\acknowledgments We thank Jon Arons, Ray Carlberg, Ron Hellings,
Yuri Levin, Andrea Lommen, John Maggorian, Sterl Phinney, Philip Stark 
and members of the
Center for Particle Astrophysics for helpful conversations.  AHJ
acknowledges support from NSF KDI grant 9872979 and NASA LTSA grant
NAG5-6552, and PPARC in the UK. DCB acknowledges support from NSF
grant AST-9731106 which partially supported the Pulsar
Timing Array experiment.

\appendix\section{Appendix}\label{appendix}
In this appendix, we calculate the probability distribution of the
quantity $h_c^2(f)$ as given by equation~(\ref{eq:fullspec}). For generality,
consider some quantity 
\begin{equation}
  \label{eq:y}
  y=\int dz\; g(z) N(z)
\end{equation}
where $N(z)\; dz$ gives the probability of some event happening in
$(z,z+dz)$ (\ie, a Poisson rate), and $g(z)$ is any function. We want
to know $P(y)$, the distribution function of $y$. We start with a few
important facts from probability theory.

First, the characteristic function (or moment generating function) of a
distribution is defined as the Fourier transform of the distribution,
\ie,
  \begin{equation}
    \label{eq:mgf}
    \varphi(t) = \langle e^{ity} \rangle = \int dy \; e^{ity} P(y)\;.
  \end{equation}
This has the property that the moments of the distribution are then
given by
\begin{equation}
\label{eq:dphidt}
  \langle y^k \rangle = i^{-k} \left.\frac{d^k\varphi(t)}{dt^k}\right|_{t=0}
\end{equation}
and the ``cumulants'' or ``connected moments'' or ``semi-invariants''
($\kappa_k$, the mean, variance, skewness, kurtosis, \ldots, for
$k=1,2,3,4,\ldots$)
are given by the same expression, substituting $\ln\varphi$ for $\varphi$.

Next, we use the fact (trivial from Fourier theory) that the
characteristic function of the sum of two variables is the product of
the individual functions, and that the characteristic function of $a y$,
where $a$ is a scalar and $y$ is the random variable is $\varphi(a t)$,
where $\varphi(t)$ is the characteristic function for $y$.

Finally, we need the characteristic function for a Poisson distribution,
$P(n)=r^n e^{-r}/n!$ where $r$ is the rate. This is just
$\varphi(t)=\exp\left[r(e^{it}-1)\right]$.

Putting all of these facts together gives the characteristic function for
$y$, which is just the sum (integral) of Poisson variables with rates
$N(z)\; dz$ times scalars $g(z)$. Thus
\begin{equation}
\label{eq:phiy}
  \ln\varphi(t) = \int dz\; N(z) \left[e^{itg(z)}-1\right]\;.
\end{equation}
The Taylor series of $\ln\varphi$ around $t=0$ then gives the cumulants of
the distribution. They are just
\begin{equation}
  \label{eq:cum}
  \kappa_k = \int dz\; \left[g(z)\right]^k N(z)
\end{equation}
(Note that the mean, $k=1$, is what would be trivially expected.)

In our case, we have $N(z)\;dz\to N(z,f,M_1,M_2)\; dz\;dM_1\;dM_2$ and
$g(z)\to h_{\rm rms}^2(z,f,M_1,M_2)$ for $y=h_c^2(f)\; df/f$, giving
equation~(\ref{eq:charfun}) above.

\clearpage

\bibliography{mbh,dbpro}
\end{document}